\documentclass{article}
\usepackage[paper=letterpaper,margin=2cm]{geometry}
\usepackage{soul, color, xcolor}
\usepackage{graphicx}
\usepackage{amsmath}
\usepackage{amssymb}
\usepackage{braket}
\usepackage{bm}
\usepackage{amsfonts}
\usepackage{cite}
\usepackage[ruled,boxed,linesnumbered]{algorithm2e}
\usepackage{tcolorbox}
\usepackage[colorlinks=true,citecolor=blue]{hyperref}
\usepackage{booktabs}
\usepackage{dashrule}
\usepackage{authblk}
\usepackage{orcidlink}
\tcbuselibrary{breakable}
\tcbset{colbacktitle=yellow!10!white, colback=white, breakable, coltitle=black, before upper={\parindent10pt\noindent}, fonttitle=\bfseries}
\date{}
\title{\textbf{Security Analysis of Mode-Pairing Quantum Key Distribution with Flexible Pairing Strategy}}

\author{\begin{minipage}{0.92\textwidth}
    \centering
    \small
    Yi-Fei Lu\orcidlink{0000-0001-8212-0556}$^1$, Yang Wang\orcidlink{0000-0001-9774-1064}$^{1,\dagger}$, Yan-Yang Zhou$^1$, Yu Zhou$^1$, Xiao-Lei Jiang$^1$, Xin-Hang Li$^2$, Hai-Tao Wang$^1$, Jia-Ji Li$^1$, Chun Zhou$^1$, Hong-Wei Li$^1$, Yu-Yao Guo$^2$, Lin-Jie Zhou$^2$, and Wan-Su Bao$^{1,*}$

    $^1$Henan Key Laboratory of Quantum Information and Cryptography, IEU, Zhengzhou 450001, China

    $^2$State Key Laboratory of Advanced Optical Communication Systems and Networks,
    
    Shanghai Key Lab of Navigation and Location Services, Department of Electronic Engineering,
    
    Shanghai Jiao Tong University, Shanghai 200240, China

    $^\dagger$wy@qiclab.cn, $^*$bws@qiclab.cn
\end{minipage}}

\begin{document}
\maketitle

{\centering\section*{Abstract}}

Mode-pairing quantum key distribution (MP-QKD) is advantageous for long-distance secure communication, leveraging its simple implementation and quadratic scaling capacity. The post-measurement pairing in MP-QKD alleviates the photon-coincidence demands, which is essential for surpassing the fundamental limit to the key-rate transmission. In this work, we propose an improved decoy-state MP-QKD protocol featuring a flexible and efficient pairing strategy. We prove the security of the proposed scheme by presenting an entanglement model for decoy-state MP-QKD. The simulation results show that the secret key rate (SKR) can be enhanced among all distances. Notably, compared with the original scheme [Nature Communication 13, 3903 (2022)], the improvement of SKR is greater than 65\% within 375 km in the asymptotic case and greater than 50\% within 400 km in the finite case. And the achievable distance can be extended in the finite case, especially with a small block length. The simulation results demonstrate the high efficiency of the proposed scheme, which is expected to promote the practical applicability of MP-QKD. Furthermore, the entanglement model could provide a theoretical framework for further security and performance analysis of decoy-state MP-QKD.

\clearpage

\section{Introduction}

Quantum key distribution (QKD) \cite{bennett2014RN153,xu2020RN130,pirandola2020RN485} is now a mature technology with the advantage of information-theoretically security. Numerous significant advancements have been demonstrated in substantial improvement of transmission distance \cite{yin2016RN288,boaron2018RN286,liao2018RN512,pittaluga2021RN482,wang2022RN585,chen2022RN735,liu2023RN975}, achieving high secret key rate (SKR) \cite{yuan2018RN513,fadri2023RN1023,li2023RN987}, integration on photonic chips \cite{ma2016RN1170,sibson2017RN1190,bunandar2018RN1195,zhang2019RN1188,wei2020RN1030,semenenko2020RN1171,wang2020RN1135}, and the deployment of quantum communication networks \cite{elliott2005RN1087,stucki2011RN1086,sasaki2011RN1085,wang2014RN1217,chen2021RN519}. These advancements have been facilitated by significant developments in theoretical protocols and improvements in experimental techniques.

QKD utilizes the single-photon states as the physical carriers for generating secret keys by employing polarization, phase, or time-bin encoding. The primary barrier of QKD is the optical loss in the quantum channel, which limits the transmission distance and SKR of point-to-point QKD without quantum repeaters \cite{pirandola2009RN1300,takeoka2014RN231,pirandola2017RN103,das2021RN719}. The fundamental rate-loss limit,  Pirandola-Laureanza-Ottaviani-Banchi (PLOB) bound \cite{pirandola2017RN103} characterizes the upper bound of SKR without quantum repeaters as $\mathcal{C}(\eta) = -\log_2(1-\eta)$, where $\eta$ is the transmission efficiency between two parties in communication. To overcome this rate-distance limit, twin-field (TF) QKD \cite{lucamarini2018RN45,ma2018RN56,wang2018RN22,curty2019RN57,maeda2019RN507,cui2019RN41} was proposed based on the single-photon interference and the advantages have been demonstrated experimentally \cite{zhong2019RN52,fang2020RN55,pittaluga2021RN482,chen2021RN586,chen2022RN735,wang2022RN585,liu2023RN975,li2023RN979,zhou2023RN949,chen2024RN1108,zhou2024RN1175}. In TF-QKD, the physical carriers are the coherent states and the information is encoded in the global phase of the states, which requires the fields generated by the two distant, independent light sources to be identical. Therefore, TF-QKD proposes two critical experimental difficulties: the phase locking of two remote lasers, and the phase tracking and compensation of the whole channel. These difficulties have been overcome but they put forward high complexity and requirements. To release these requirements while maintaining the performance, mode-pairing (MP) QKD \cite{zeng2022RN816} (also called asynchronous measurement-device-independent (MDI) QKD\cite{xie2022RN724}) was proposed, which can be seen as an improved version of MDI QKD \cite{lo2012RN72,braunstein2012RN581,ma2012RN73}. In MDI-QKD, Alice and Bob prepare the paired bins at the beginning. The raw key bits can only be generated when two bins are detected coincidentally, which limits the SKR scaling to $O(\eta)$. In MP-QKD, it does not assume two pre-determined bins as a coding pair, but instead pairs the bins based on the results announced by Charlie. The flexibility of the pairing leads to significant improvements in the efficiency of MP-QKD with SKR scaling to $O(\sqrt{\eta})$ \cite{zeng2022RN816,xie2022RN724,lu2025RN1293}. Since the information in MP-QKD is encoded in the relationship between two bins that travel through the same quantum channel, the protocol is inherently insensitive to phase drift. Besides, the requirement of the wavelength consistency of two remote lasers in MP-QKD is still higher but is relatively lower than in TF-QKD. At present, the efficiency and simplicity of MP-QKD have been successfully demonstrated \cite{zhu2023RN928,zhou2023RN978,zhu2024RN1180,ge2025RN1181}. 

The pairing strategy is crucial for both the security and efficiency of MP-QKD. It is necessary to design a secure and more efficient pairing strategy. First, the paired bins can not be sifted in too large a time interval due to the limited coherence time of the lasers and phase drift in the quantum channel. The pairing interval is set to avoid these impacts \cite{zeng2022RN816,xie2022RN724,zhu2023RN928,zhou2023RN978,zhu2024RN1180,ge2025RN1181}. The thorough optimization analysis for the maximum pairing length is presented by considering practical conditions \cite{zhou2025RN1272}. Second, not all pairing strategy is permitted to ensure the independence of the paired states and the fairness of the parameter estimation \cite{zeng2022RN816,lu2025RN1293}. Third, the SKR may vary with different pairing strategies. The simple pairing strategy is to pair two adjacent successfully detected bins within the pairing interval \cite{zeng2022RN816}. In this case, the pairing is only determined by Charlie's announced measurement results and the security can be guaranteed \cite{zeng2022RN816}. However, the efficiency is limited as partial pairs (e.g., when signal and decoy states are paired) are useless when the decoy-state method is applied. To reduce the number of useless pairs, a scheme is designed by announcing the decoy state after Charlie's announcement to filter out the rounds that are useless for pairing \cite{zhou2023RN978,xie2023RN1097}. From another perspective, a re-pairing strategy is proposed to improve the pairing efficiency of X basis to guarantee better parameter estimation \cite{zhou2025RN1275}. In addition, many practical issues, e.g., finite effect \cite{wang2023RN1051}, asymmetric setting \cite{lu2024RN1183,li2024RN1225}, intensity fluctuation \cite{li2024RN1271}, have been analyzed to promote its practicality. The advantage distillation method \cite{liu2023RN1182} and the wavelength-division multiplexing scheme \cite{zhou2024RN1084} for MP-QKD have been proposed to improve its performance.

In this work, we present an entanglement model for optimizing and analyzing the security of the decoy-state MP-QKD. By employing this model, we design a flexible and effective pairing strategy for decoy-state MP-QKD to improve the transmission distance and SKR. We prove rigorously the security of decoy-state MP-QKD featuring a flexible pairing strategy. The proposed pairing strategy enhances the efficiency of raw key generation by improving the pairing efficiency of the Z basis. In Sec. \ref{protocol}, we present the decoy-state MP-QKD protocol and the flexible pairing strategy. We introduce the entanglement model for decoy-state MP-QKD and give the detailed parameter estimation method with flexible pairing strategy in Sec. \ref{security_analysis}. In Sec. \ref{performance} we show the simulation method of SKR to compare the flexible pairing strategy with the original pairing strategy \cite{zeng2022RN816}. The results show that the SKR is improved under the practical parameters \cite{zhou2023RN949} at all distances. In the asymptotic case, the improvement is generally greater than 65\%. In the finite case, the improvement is greater than 50\% and could be enhanced as the distance increases with small block lengths. Moreover, the achievable distance can be expanded especially with a small block length. Lastly, the conclusion is given in Sec. \ref{conclusion}.

\section{Protocol}
\label{protocol}

\begin{figure}[t]
    \centering
    \includegraphics[width=0.91\textwidth]{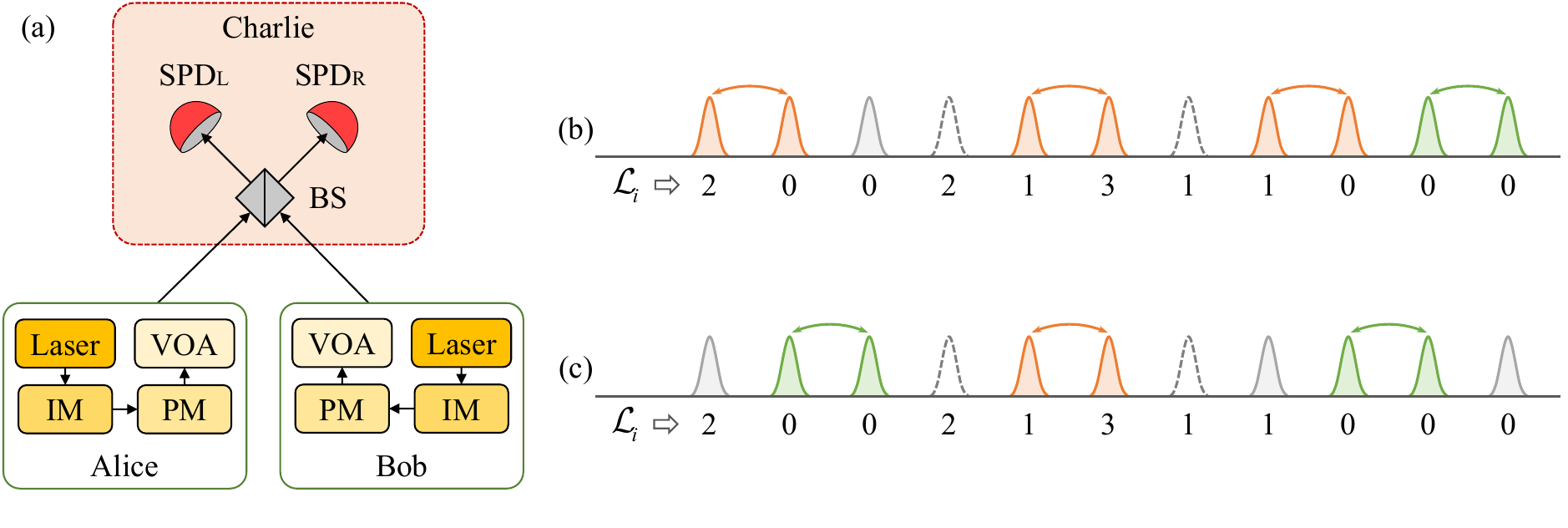}
    \caption{Schematics of MP-QKD protocol and pairing strategies. (a) Sketch of MP-QKD protocol. Alice and Bob employ the continuous-wave lasers as the sources. The phase randomized weak coherent pulses are prepared with the intensity modulator (IM), phase modulator (PM) and variable optical attenuator (VOA). Charlie performs the interference measurement with a 50:50 beam splitter (BS) and two single-photon detectors (SPDs). (b) An example of the pairing results is based on the original pairing strategy, which pairs adjacent effective rounds directly. Here we assume the maximal pairing interval $l = 2$ for simplicity. The solid and dashed pulses correspond to the effective and ineffective rounds, respectively. The gray solid pulses denote the unpaired rounds, and the green and red solid pulses denote the paired rounds. Those green pairs can be used to generate raw key bits, and the red pairs are useful for parameter estimation. (c) An example of the pairing results based on the pairing strategy in Box 2. The rounds with label $\mathcal{L}_i = 2$ are first filtered out, and the rounds with label $\mathcal{L}_i = 1$ or 3 are filtered out with probability $1 - p_{\text{save}}$. Then the saved and adjacent rounds are paired within the maximal pairing interval $l = 2$.}
    \label{fig_mp_system}
\end{figure}

In this section, we present the vacuum + weak decoy-state MP-QKD protocol in Box 1 equipped with the flexible pairing strategy in Box 2. In every round, Alice and Bob will independently prepare the phase-randomized coherent states with three different intensities, one of which is the vacuum state. Then the prepared states are sent to the third party Charlie for detection. After the detection, Alice and Bob will determine the labels for every round according to the intensities. Based on Charlie's announced detection results and the labels, Alice and Bob can perform the pairing strategy in Box 2 and extract the secret keys in the post-processing steps. A sketch of MP-QKD protocol is shown in Fig. \ref{fig_mp_system}(a). Below we present the detailed description of the scheme and provide the notations.

\begin{tcolorbox}[title = {Box 1: Decoy-State MP-QKD}]

    \textbf{(1) State preparation}. In the $k$-th round, Alice prepares a coherent state $\ket{e^{i\theta_k^a} \sqrt{\tau_k^a}}$, where the random phase $\theta_k^a \in [0,2\pi)$ and the intensity $\tau_k^a$ is chosen from $\{\mu_a,\nu_a,o_a\}$ with probability $p_{\tau_k^a}$. Here, $o_a = 0$ denotes the vacuum states. Alice records the local bits $a_k$ as 0, 1 or 2 when $\tau_k^a = o_a$, $\mu_a$ or $\nu_a$. Bob performs the same procedure independently and his notations can be defined similarly by changing the indices $a$ to $b$.

    \textbf{(2) Measurement}. Alice and Bob send the prepared states to Charlie, who is supposed to perform the interference measurement and announce the results $L_k,R_k \in \mathbb{Z}_2$ of two single-photon detectors (SPDs). The results 1 and 0 denote whether Charlie announced an SPD click or not.

    \textbf{(3) Mode pairing}. After repeating the above steps $N$ times, Alice and Bob sift the effective rounds when $L_k \oplus R_k = 1$. For those effective rounds, they announce the round when $a_k$ or $b_k$ is 2 and then announce the corresponding $b_k$ or $a_k$. They assign the label $\mathcal{L}_k$ as 0 when the round is not announced, as 1 when $(a_k|b_k)=(0|2)$ or $(2|0)$, as 2 when $(a_k|b_k)=(1|2)$ or $(2|1)$, or as 3 when $(a_k|b_k)=(2|2)$. Then they perform the pairing according to the pairing strategy in Box 2.

    \textbf{(4) Basis sifting}. Alice assigns the pairs (indexed by $j$ and $k$) as $Z$ basis if $(a_j,a_k)=(0,1)$ or $(1,0)$, as $X$ basis if $(a_j,a_k) = (2,2)$, and as '0' if $(a_j,a_k) = (0,0)$. Bob assigns the basis independently according to $b_j$ and $b_k$. Then they announce the basis and sift the pairs when both are $Z$ basis to perform key mapping. Other pairs can be used to perform the decoy-state method. Denote the number of pairs corresponding to intensities $\tau_a = \tau_j^a +\tau_k^a$ and $\tau_b = \tau_j^b +\tau_k^b$ as $n_{[\tau_a,\tau_b]}$, apart from $n_{[2\nu_a,2\nu_b]}$ which is defined in the next step.

    \textbf{(5) Key mapping}. In $Z$ basis, Alice assigns the raw key bits as $a_j$ and Bob conversely assigns the raw key bits as $b_k$. When both are $X$ basis, Alice and Bob announce the phase differences $\delta_{jk}^a = (\theta_j^a - \theta_k^a) \text{mod} 2\pi$ and $\delta_{jk}^b = (\theta_j^b - \theta_k^b) \text{mod} 2\pi$ and sift the pairs satisfying $|\delta_{jk}^a - \delta_{jk}^b| < 2\pi / M$ or $||\delta_{jk}^a - \delta_{jk}^b| - \pi| < 2\pi / M$ to estimate the phase error rate. The error pairs are defined in two cases: $L_j = L_k$ when $||\delta_{jk}^a - \delta_{jk}^b| - \pi| < 2\pi / M$, $L_j \neq R_k$ when $|\delta_{jk}^a - \delta_{jk}^b| < 2\pi / M$. Here, $M$ is the number of phase slices and should be set properly (e.g., $M=16$) to balance the error rate and the number of pairs \cite{lucamarini2018RN45,ma2012RN73}. Besides, the discrete-phase-randomized phase modulation is often adopted in practical application \cite{cao2015RN105,zhang2024RN1294}. Define the number of the sifted pairs and error pairs as $n_{[2\nu_a,2\nu_b]}$ and $m_{[2\nu_a,2\nu_b]}$, respectively.

    \textbf{(6) Parameter estimation}. By employing the decoy-state method, they estimate the bounds of the number of bits $n_{11}^z$ and the phase error rate $e_{11}^x$ when both Alice and Bob send the single-photon state $(\ket{10}\bra{10} + \ket{01}\bra{01})/2$ in $Z$ basis. The details for the parameter estimation are presented in Sec. \ref{sec_parameter_estimation}.

    \textbf{(7) Key distillation}. Alice and Bob perform the error correction and privacy amplification steps to distill the final key bits. The SKR in the finite case is given by \cite{zhou2023RN978,zeng2022RN816,xie2023RN1097}
    \begin{equation}
        \label{eq4}
        R = \frac{1}{N} \bigg\{\underline{n}_{11}^z [1 - H_2(\overline{e}_{11}^x)] - \lambda_{\text{EC}} - \log_2 \frac{2}{\varepsilon_{\text{cor}}} - 2\log_2 \frac{2}{\varepsilon^\prime \hat{\varepsilon}} - 2\log_2 \frac{1}{2\varepsilon_{\text{PA}}} \bigg\},
    \end{equation}
    where $\underline{x}$ and $\overline{x}$ denote the lower and upper bounds of a parameter $x$, $H_2(x) = -x\log_2(x) - (1-x) \log_2(1-x)$ is the binary Shannon entropy function, $\lambda_{\text{EC}}$ is the information revealed in the error correction step, and $\varepsilon_{\text{cor}}, \varepsilon^\prime, \hat{\varepsilon}, \varepsilon_{\text{PA}}$ are the security coefficients regarding the correctness and secrecy.

\end{tcolorbox}

Below we analyze and present the flexible pairing strategy. The primary aim is to improve the efficiency of $Z$ pairs. However, the SKR may be low in the finite case due to the statistical fluctuation of other pairs for parameter estimation. Therefore, we should balance the number of different pairs in the finite case to obtain the optimal SKR. First, some rounds are useless for both key generation and parameter estimation and should be discarded promptly. Second, those useful rounds may form useless pairs, which should be suppressed. Third, the number of different pairs should be balanced flexibly to obtain the optimal SKR. Fourth, the optimization of the balance should be simple to perform. Based on these considerations, we show the pairing strategy in Box 2. The pairing strategy can be divided into two stages. In the first stage, Alice and Bob assign the auxiliary parameter $C_i^\prime$ to replace $C_i$ for every round. Then they perform pairing in the second stage according to parameter $C_i^\prime$ the same as that in the original pairing strategy \cite{zeng2022RN816}. The key and flexibility of the pairing strategy lie in the assignment of the parameter $C_i^\prime$. If the label $\mathcal{L}_i=0$, then $C_i^\prime$ is assigned as $C_i$. These rounds can be used to form $Z$ pairs. If the label $\mathcal{L}_i=2$, then $C_i^\prime$ is assigned as 0, which is useless. For other rounds, $C_i^\prime$ is assigned as $C_i$ and 0 with probabilities $p_{\text{save}}$ and $1 - p_{\text{save}}$, respectively. These rounds can form the pairs for parameter estimation. Note these rounds will affect the pairing of $Z$ pairs, e.g., if the two rounds with $(\mathcal{L}_j,\mathcal{L}_k) = (0,1)$ or $(1,0)$ are adjacent in the pairing interval. We could optimize the probability $p_{\text{save}}$ flexibly to obtain the optimal SKR.

\begin{tcolorbox}[title = {Box 2: Pairing Strategy}]

    \textbf{Input}: The label $\mathcal{L}_i$, Charlie's announced detection results $C_i = L_i \oplus R_i$ for $i=1$ to $N$; pairing interval $l$; $p_{\text{save}} \in (0,1]$.

    \textbf{Output}: $K$ pairs, $(F_k, R_k)$ as the $k$-th pair for $k=1$ to $K$.

    \textbf{Initialization}: $k=1$, $f=0$; $C_i^\prime = C_i$, $s_i=0$ and 1 with probabilities $1-p_{\text{save}}$ and $p_{\text{save}}$ for $i=1$ to $N$.

    \vspace{-0.5em}
    \noindent\hdashrule{\textwidth}{0.6pt}{4pt 2.5pt}

    \textbf{for} $i=1$ to $N$ \textbf{do} \hfill $\rhd$ Enumerating all locations

    \quad \textbf{if} $\mathcal{L}_i = 0$ \textbf{then}

    \quad\quad $C_i^\prime \leftarrow C_i$. \hfill $\rhd$ Keeping this round

    \quad\textbf{else if} $\mathcal{L}_i = 2$ \textbf{then}

    \quad\quad $C_i^\prime \leftarrow 0$. \hfill $\rhd$ Discarding this round

    \quad\textbf{else}

    \quad\quad $C_i^\prime \leftarrow s_i C_i$. \hfill $\rhd$ Keeping this round with practicality $p_{\text{save}}$

    \quad \textbf{end if}

    \textbf{end for}

    \vspace{-0.5em}
    \noindent\hdashrule{\textwidth}{0.6pt}{4pt 2.5pt}

    \textbf{for} $i=1$ to $N$ \textbf{do} \hfill $\rhd$ Enumerating all locations

    \quad \textbf{if} $C_i^\prime = 1$ \textbf{then} \hfill $\rhd$ Searching effective and kept round

    \quad\quad \textbf{if} $f=0$ \textbf{then} \hfill $\rhd$ Searching for the front-pulse location
    
    \quad\quad\quad $F_k \leftarrow i$; $f \leftarrow 1$.

    \quad\quad \textbf{else then} \hfill $\rhd$ Searching for the rear-pulse location

    \quad\quad\quad \textbf{if} $i - F_k \leq l$ \textbf{then} \hfill $\rhd$ Pairing interval less than $l$

    \quad\quad\quad\quad $R_k \leftarrow i$; $k \leftarrow k+1$; $f \leftarrow 0$.

    \quad\quad\quad \textbf{else then} \hfill $\rhd$ Pairing interval exceeding $l$
    
    \quad\quad\quad\quad $F_k \leftarrow i$.

    \quad\quad\quad \textbf{end if}

    \quad\quad \textbf{end if}

    \quad \textbf{end if}

    \textbf{end for}

    Set the total number of pairs $K = k-1$.

\end{tcolorbox}

According to the pairing strategy in Box 2, each effective round has a certain probability of being saved for pairing, which is different from the original pairing strategy. We define $p_{\text{save}}^{(\tau_k^a,\tau_k^b)}$ as the probability that an effective round with Alice and Bob's intensities $\tau_k^a$ and $\tau_k^b$ can be saved for pairing. The probability $p_{\text{save}}^{(\tau_k^a,\tau_k^b)}$ may take three values $\{0, p_{\text{save}}, 1\}$. The label $\mathcal{L}_k$ is directly determined by the intensities $\tau_k^a$ and $\tau_k^b$ according to step (3) in Box 1. Therefore, the probability $p_{\text{save}}^{(\tau_k^a,\tau_k^b)}$ can be obtained as
\begin{equation}
    \begin{aligned}
        \label{eq8}
        p_{\text{save}}^{(\tau_k^a,\tau_k^b)} =
        \begin{cases}
        1,               & \text{if} \ \mathcal{L}_k = 0, \\
        0,               & \text{if} \ \mathcal{L}_k = 2, \\
        p_{\text{save}}, & \text{if} \ \mathcal{L}_k = 1 \ \text{or} \ 3.
        \end{cases}
    \end{aligned}
\end{equation}

In Fig. \ref{fig_mp_system}(b) and (c), we show the simple examples of the pairing results to compare the pairing strategy in Box 2 and the original pairing strategy \cite{zeng2022RN816}. In Fig. \ref{fig_mp_system}(b), the pairing results are obtained by pairing to adjacent effective rounds without referring to the labels $\mathcal{L}_i$. In Fig. \ref{fig_mp_system}(c), the pairing process first filters out partial rounds that are ineffective for raw key bits based on the labels $\mathcal{L}_i$, and then pairs the remaining adjacent effective rounds. The number of all pairs in Fig. \ref{fig_mp_system}(c) is less than that in Fig. \ref{fig_mp_system}(b) but the $Z$ pairs for raw key bits are more. Therefore, this pairing strategy in Box 2 would help to improve the SKR.

\section{Security analysis}
\label{security_analysis}

In this section, we present the entanglement model for the decoy-state MP-QKD equipped with the flexible pairing strategy. Besides, we provide a detailed parameter estimation method for the SKR in the asymptotic and finite scenarios.

\subsection{Entanglement model}

To give the entanglement scheme, we propose an entanglement scheme for the decoy-state MP-QKD in Box 3. We prove the security of the entanglement scheme and then prove the equivalence between the entanglement scheme in Box 3 and the prepare-and-measure scheme in Box 1. In this way, we can perform the security and performance analysis directly on the entanglement scheme to obtain the properties of the prepare-and-measure scheme. The entanglement model provides a theoretical framework for the analysis of decoy-state MP-QKD.

We first analyze how to give the entanglement scheme. In step (1) of Box 1, Alice will prepare three coherent states with different probabilities, which can be expressed as the mixed state in the actual system $A_k$
\begin{equation}
    p_{o_a} \ket{o_a}\bra{o_a} + p_{\mu_a} \ket{e^{i\theta_k^a}\mu_a}\bra{e^{i\theta_k^a}\mu_a} + p_{\nu_a} \ket{e^{i\theta_k^a}\nu_a}\bra{e^{i\theta_k^a}\nu_a}.
\end{equation}
We could introduce an ancillary system $A_k^\prime$ to purify the mixed states as
\begin{equation}
    \label{eq7}
    \ket{\varphi}_{A_k^\prime A_k} = \big(\sqrt{p_{o_a}} \ket{0}\ket{o_a} + \sqrt{p_{\mu_a}} \ket{1}\ket{e^{i\theta_k^a}\sqrt{\mu_a}} + \sqrt{p_{\nu_a}} \ket{2}\ket{e^{i\theta_k^a}\sqrt{\nu_a}} \big)_{A_k^\prime A_k}.
\end{equation}
Note that this purification is virtual and local, which will not affect the actual procedure. Below we start from this state to give the entanglement scheme to show how to establish entanglement between Alice and Bob by introducing local measurements.

\begin{tcolorbox}[title = {Box 3: Entanglement Scheme for Decoy-State MP-QKD}]

    \textbf{(i) State preparation}. In the $k$-th round, Alice chooses a random phase $\theta_k^a \in [0,2\pi)$ and prepares the composite states $\ket{\varphi}_{A_k^\prime A_k}$. Here, the systems $A_k^\prime$ and $A_k$ are the ancillary and practical systems, respectively. Bob performs the same procedure independently.

    \textbf{(ii) Measurement}. Alice and Bob send the systems $A_k$ and $B_k$ to Charlie, and Charlie performs the same as step (2) in Box 1.

    \textbf{(iii) Mode pairing}. After repeating the above steps $N$ times, Alice and Bob sift the effective rounds with $L_k \oplus R_k = 1$. Then they perform the collective measurement on the systems $A_k^\prime B_k^\prime$ in the effective rounds with operators
    \begin{equation}
        \begin{aligned}
            M_{0} &= (\ket{0}\bra{0} + \ket{1}\bra{1}) \otimes (\ket{0}\bra{0} + \ket{1}\bra{1}),\\
            M_{1} &= \ket{20}\bra{20} + \ket{02}\bra{02},\\
            M_{2} &= \ket{21}\bra{21} + \ket{12}\bra{12},\\
            M_{3} &= \ket{22}\bra{22}.
        \end{aligned}
    \end{equation}
    and obtain the measurement result as the label $\mathcal{L}_k \in \mathbb{Z}_4$. Note that the measurement satisfies the completeness relation as $\sum_{i=0}^{3} M_i^\dagger M_i = \mathbb{I}$. They perform the pairing according to the pairing strategy in Box 2.

    \textbf{(iv) Basis sifting}. For the pairs, Alice (Bob) measures the composite systems $A_j^\prime A_k^\prime$ ($B_j^\prime B_k^\prime$) with the measurement operators
    \begin{equation}
        \begin{aligned}
            M_0^\prime &= \ket{00}\bra{00},\\
            M_1^\prime &= \ket{01}\bra{01} + \ket{10}\bra{10},\\
            M_2^\prime &= \ket{22}\bra{22},\\
            M_3^\prime &= \mathbb{I} - M_0 - M_1 - M_2,
        \end{aligned}
    \end{equation}
    and assigns the pair as $Z$ when the measurement result is 1, as $X$ basis when 2, and as '0' when 0. Others are the same as step (4) in Box 1.

    \textbf{(v) Key mapping}. Alice and Bob announce the phase differences $\delta_{jk}^a$ and $\delta_{jk}^b$ of the pairs. When both Alice and Bob are $Z$ basis, Bob evolves the ancillary systems $B_j^\prime B_k^\prime$ with the operator $U_{\delta_{jk}^a-\delta_{jk}^b}$, which $U_\delta$ is a phase gate and defined as
    \begin{equation}
        U_\delta = \ket{01}\bra{01} + e^{i\delta} \ket{10}\bra{10}.
    \end{equation}
    Then Alice (Bob) measures the composite systems $A_j^\prime A_k^\prime$ ($B_j^\prime B_k^\prime$) in the basis $\{\ket{xy}\}_{x,y\in\mathbb{Z}_3}$ and obtains the results $a_j,a_k$ ($b_j, b_k$) in $Z$ basis. Others are the same as step (5) in Box 1. 

    \textbf{(vi)-(vii)}. Same as step (6)-(7) in Box 1.

\end{tcolorbox}

In step (i), Alice and Bob will prepare the composite states $\ket{\varphi}_{A_k^\prime A_k} \ket{\varphi}_{B_k^\prime B_k}$ in the $k$-th round. Then they will perform the collective measurement on the ancillary systems $A_k^\prime B_k^\prime$ and assign the label $\mathcal{L}_k$ in step (iii). When the label $\mathcal{L}_k = 0$, the (unnormalized) result state is
\begin{equation}
    \begin{aligned}
        \rho_0 = \sum_{\chi \in \mathbb{Z}_2^2} \hat{P} \big[ & E_\chi (\sqrt{p_{o_a}} \ket{0}\ket{o_a} + \sqrt{p_{\mu_a}} \ket{1}\ket{e^{i\theta_k^a}\sqrt{\mu_a}})_{A_k^\prime A_k}  \otimes (\sqrt{p_{o_b}} \ket{0}\ket{o_b} + \sqrt{p_{\mu_b}} \ket{1}\ket{e^{i\theta_k^b}\sqrt{\mu_b}})_{B_k^\prime B_k} \big],
    \end{aligned}
\end{equation}
where $\hat{P}(\ket{x}) = \ket{x}\bra{x}$, and $E_\chi$ is the operator on the systems $A_kB_k$ corresponding to announced measurement results $\chi = (L_k, R_k)$. The cases with other labels will be analyzed when needed. In step (iv), Alice and Bob will independently measure the ancillary systems $A_j^\prime A_k^\prime$ and $B_j^\prime B_k^\prime$ of the pairs. If the labels $\mathcal{L}_j$ and $\mathcal{L}_k$ are both 0 in the paired rounds and the measurement results are both 1 in step (iv), the result state is
\begin{equation}
    \sigma_{1} = \sum_{\chi^\prime \in \mathbb{Z}_2^4} \hat{P} \big( E_{\chi^\prime} \ket{\varphi_{\mu_a,\theta_j^a,\theta_k^a}}_{A_j^\prime A_k^\prime A_j A_k} \ket{\varphi_{\mu_b,\theta_j^b,\theta_k^b}}_{B_j^\prime B_k^\prime B_j B_k} \big),
\end{equation}
where $E_{\chi^\prime}$ is the operator on the systems $A_jB_j$ and $A_kB_k$ corresponding to announced measurement results $\chi^\prime = (L_j, R_j,L_k, R_k)$, and the state is defined by eliminating the subscripts of the intensities as
\begin{equation}
    \ket{\varphi_{\mu,\theta_1,\theta_2}} = \frac{1}{\sqrt{2}} \big(\ket{10} \ket{e^{i\theta_1}\sqrt{\mu}} \ket{o} + \ket{01} \ket{o} \ket{e^{i\theta_2}\sqrt{\mu}}\big).
\end{equation}
In step (v), Alice and Bob will only announce the phase differences $\delta_{jk}^a$ and $\delta_{jk}^b$ but not the specific phases. Hence, the states $\ket{\varphi_{\mu,\theta_1,\theta_2}}$ is equivalent to the following state
\begin{equation}
    \rho_{\mu} = \sum_{s\in\mathbb{N}} p_{s,\mu} \hat{P} (\ket{\varphi_{s,\delta}}),
\end{equation}
where the Poisson distribution probability is defined as
\begin{equation}
    p_{s,\tau} = e^{-\tau} \frac{\tau^s}{s!},
\end{equation}
the phase difference $\delta = (\theta_1 - \theta_2) \text{mod} 2\pi$, and the state $\ket{\varphi_{s,\delta}}$ is defined as
\begin{equation}
    \label{eq6}
    \ket{\varphi_{s,\delta}} = \frac{1}{\sqrt{2}} (\ket{01}\ket{0s} + e^{i\delta}\ket{10}\ket{s0} ).
\end{equation}
It is equivalent that Alice (Bob) has prepared the state $\rho_{\mu}$ when the labels are 0 in the paired rounds and the measurement in step (iv) is 1. The single-photon part in $\rho_{\mu}$ is $\ket{\varphi_{1,\delta}} = (\ket{01}\ket{01} + e^{i\delta}\ket{10}\ket{10} )/ \sqrt{2}$, which can be used to distill secret keys. Besides, Bob will evolve the phase in $Z$ basis in step (v) to guarantee the single-photon states for them are both $\ket{\varphi_{1,\delta}}$ with random phase $\delta = \delta_{jk}^a$. Then they measure the ancillary systems to obtain the results and assign the raw key bits. By applying the tagging method \cite{gottesman2004RN266}, the SKR can be obtained in Eq. (\ref{eq4}).

The phase error rate in Eq. (\ref{eq4}) characterizes the information leakage. To analyze the phase error rate, we define the two-mode $s$-photon states $\ket{\omega_{s,\delta}}$ as
\begin{equation}
    \ket{\omega_{s,\delta}} = \frac{1}{\sqrt{2^s}} \sum_{r=0}^s \sqrt{C_s^r} e^{ir\delta} \ket{r} \ket{s-r}.
\end{equation}
The single-photon entanglement state $\ket{\varphi_{1,\delta}}$ in Eq. (\ref{eq6}) can be reformulated as
\begin{equation}
    \ket{\varphi_{1,\delta}} = \frac{1}{\sqrt{2}} (\ket{\omega_{1,0}} \ket{\omega_{1,\delta}} + \ket{\omega_{1,\pi}} \ket{\omega_{1,\delta + \pi}}).
\end{equation}
If Alice and Bob have virtually measured the state in basis $\{\ket{\omega_{1,0}}, \ket{\omega_{1,\pi}}\}$ in step (v), the emitted states are $\ket{\omega_{1,\delta}}$ and $\ket{\omega_{1,\delta + \pi}}$, respectively. The phase error rate $e_{11}^x$ is defined as the fraction of two cases: $L_j \neq L_k$ when Alice and Bob emit the same states; $L_j = L_k$ when Alice and Bob emit the different states. Below we analyze how to estimate the phase error rate.

When the measurement result $\mathcal{L}_k = 3$ in step (iii), the result state is
\begin{equation}
    \rho_2 = \sum_{\chi \in \mathbb{Z}_2^2} \hat{P} \big( E_\chi \ket{2}_{A_k^\prime} \ket{e^{i\theta_k^a}\sqrt{\nu_a}}_{A_k} \ket{2}_{B_k^\prime} \ket{e^{i\theta_k^b}\sqrt{\nu_b}}_{B_k} \big).
\end{equation}
The pairs with labels $\mathcal{L}_j = \mathcal{L}_k = 2$ are $\rho_2 \otimes \rho_2$. The measurement results in step (iv) will be 2 and the states are unchanged, which can be reformulated as
\begin{equation}
    \sigma_{2} = \sum_{\chi^\prime \in \mathbb{Z}_2^4} \hat{P} \big( E_{\chi^\prime} \ket{\varphi_{2\nu_a,\theta_j^a,\theta_k^a}}_{A_j^\prime A_k^\prime A_j A_k} \ket{\varphi_{2\nu_b,\theta_j^b,\theta_k^b}}_{B_j^\prime B_k^\prime B_j B_k} \big).
\end{equation}
where
\begin{equation}
    \ket{\varphi_{2\nu,\theta_1,\theta_2}} = \ket{22} \ket{e^{i\theta_1}\sqrt{\nu}} \ket{e^{i\theta_2}\sqrt{\nu}}.
\end{equation}
After the announcement of the phase differences, the state $\ket{\varphi_{2\nu,\theta_1,\theta_2}}$ is equivalent to the following state
\begin{equation}
    \rho_{2\nu} = \sum_{s\in\mathbb{N}} p_{s,2\nu} \hat{P} (\ket{22} \ket{\omega_{s,\delta}}),
\end{equation}
The single-photon part in $\rho_{2\nu}$ is $\ket{22} \ket{\omega_{1,\delta}}$, which can be used to estimate the phase error rate. By sifting pairs satisfying $\delta_{jk}^a = \delta_{jk}^b \text{mod} \pi$ and performing the decoy-state method, we can estimate the phase error rate as discussed in Sec. \ref{sec_parameter_estimation}. In practical systems, we could add the phase slices to improve the number of pairs as discussed in step (5).

In this way, we show the security of the entanglement scheme for decoy-state MP-QKD in Box 3. Below we prove the equivalence of the entanglement scheme and the prepare-and-measure scheme in Box 1. The key is to move the additional operation in the entanglement scheme (i.e., the measurements in step (iii)-(v), and the phase evolution in step (v)) to step (i) and reduce it to the prepare-and-measure scheme. The difficulty is that these operations are collective and conditioned on Charlie's announced detection results.

In the entanglement scheme, Charlie only participates in the protocol in step (ii), and other steps are performed only by Alice and Bob. The measurement in step (iii) is to distinguish the kinds of rounds. The measurement in step (iv) is to distinguish the kinds of pairs. And the operator $U_\delta$ in step (v) is used to evolve the phase. But the raw key bits only are determined until the last measurement in step (v). That is to say, the raw key bits are only determined at the end of the quantum stage. Based on this, we try to combine these quantum operations.

First, the phase evolution with phase gate $U_\delta$ is only performed in $Z$ basis, and it has no physical observability on the subsequent measurement in the basis $\{\ket{xy}\}_{x,y \in \mathbb{Z}_3}$. If this phase evolution is eliminated, the key extraction will not be affected. Besides, the estimation of the phase error rate with $X$ basis will not be affected as the phase evolution is only performed in $Z$ basis. If we combine the measurements in steps (iv) and (v) as one measurement, the last measurement results $a_j,a_k,b_j,b_k$ in step (v) can still be obtained. Besides, we could infer the measurement results in step (iv) based on the measurement results in step (v) and could assign the pairs the same as step (4) in Box 1. The composite measurement is just the measurement in step (v). In this way, the entanglement scheme in Box 3 is equivalent to the following scheme in Box 4 where Alice and Bob perform the composite measurement in step (iv').

\begin{tcolorbox}[title = {Box 4: Entanglement Scheme for Decoy-State MP-QKD II}]

    \textbf{(i')-(iii')}. Same as step (i)-(iii) in Box 3.

    \textbf{(iv') Basis sifting}. For the pairs, Alice (Bob) measures the composite systems $A_j^\prime A_k^\prime$ ($B_j^\prime B_k^\prime$) in the basis $\{\ket{xy}\}_{x,y\in\mathbb{Z}_3}$ and obtain the results $a_j,a_k$ ($b_j, b_k$). Others are the same as step (4) in Box 1.

    \textbf{(v')-(vii')}. Same as step (5)-(7) in Box 1.

\end{tcolorbox}

In Box 4, the measurement on the systems $A_j^\prime A_k^\prime$ ($B_j^\prime B_k^\prime$) in the basis $\{\ket{xy}\}_{x,y\in\mathbb{Z}_3}$ is equivalent to the measurement on the systems $A_j^\prime$ and $A_k^\prime$ ($B_j^\prime$ and $B_k^\prime$) in the basis $\{\ket{x}\}_{x\in\mathbb{Z}_3}$, separately. Therefore, Alice (Bob) could measure the systems $A_j^\prime$ and $A_k^\prime$ separately in the basis $\{\ket{x}\}_{x\in\mathbb{Z}_3}$ to obtain the results $a_j$ and $a_k$. This gives the following scheme in Box 5, which is equivalent to the entanglement scheme in Box 4.

\begin{tcolorbox}[title = {Box 5: Entanglement Scheme for Decoy-State MP-QKD III}]

    \textbf{(i'')-(iii'')}. Same as step (i)-(iii) in Box 3.

    \textbf{(iv'') Basis sifting}. Alice (Bob) measures the composite system $A_k^\prime$ ($B_k^\prime$) in the basis $\{\ket{x}\}_{x\in\mathbb{Z}_3}$ and obtains the results $a_k$ ($b_k$). Others are the same as step (4) in Box 1.

    \textbf{(v'')-(vii'')}. Same as step (5)-(7) in Box 1.

\end{tcolorbox}

In Box 5, there are now two measurements in steps (iii'') and (iv''). The measurement results in step (iii'') can still be inferred from the measurement results in step (iv''), which can be inferred as step (3) in Box 1. Hence we try to combine these two measurements as one. And the composite measurement is just the measurement in step (iv''). Hence the entanglement scheme in Box 5 is equivalent to the following scheme in Box 6.

\begin{tcolorbox}[title = {Box 6: Entanglement Scheme for Decoy-State MP-QKD IV}]

    \textbf{(i''')-(ii''')}. Same as step (i)-(ii) in Box 3.

    \textbf{(iii''') Mode pairing}.  Alice (Bob) measures the system $A_k^\prime$ ($B_k^\prime$) in the basis $\{\ket{x}\}_{x\in\mathbb{Z}_3}$ and obtain the results $a_k$ ($b_k$). Others are the same as step (3) in Box 1.

    \textbf{(iv''')-(vii''')}. Same as step (4)-(7) in Box 1.

\end{tcolorbox}

The measurement on the ancillary systems in step (iii''') is performed on the ancillary systems in every round, which is not based on Charlie's operators. Therefore, this measurement is commuted with Eve's evolution on the practical systems. Thus, Alice and Bob could advance the measurement in the state-preparation step, which is present in the following protocol.

\begin{tcolorbox}[title = {Box 7: Entanglement Scheme for Decoy-State MP-QKD V}]

    \textbf{(1') State preparation}. In the $k$-th round, Alice chooses a random phase $\theta_k^a \in [0,2\pi)$ and prepares the composite states $\ket{\varphi}_{A_k^\prime A_k}$. Alice measures the ancillary system $A_k^\prime$ in the basis $\{\ket{x}\}_{x\in\mathbb{Z}_3}$ and obtain the result $a_k$. Bob performs the same procedure independently.

    \textbf{(2')-(7')}. Same as step (2)-(7) in Box 1.

\end{tcolorbox}

The preparation of the ancillary $A_k^\prime$ and $B_k^\prime$ can be removed and the entanglement scheme in Box 7 is equivalent to the prepare-and-measure scheme in Box 1. In this way, the prepare-and-measure scheme MP-QKD in Box 1 is equivalent to the entanglement scheme MP-QKD in Box 3, which proves the security of the former.

\subsection{Parameter estimation method}
\label{sec_parameter_estimation}

To obtain the SKR, we should estimate the parameters about the single-photon states, i.e., the lower bound of $n_{11}^z$ and upper bound of $e_{11}^x$ in Eq. (\ref{eq4}). Below we analyze how to obtain these bounds with the decoy-state method. 

The number of the pairs $n_{[\tau_a,\tau_b]}$, which is defined in step (4) of Box 1, can be decomposed as
\begin{equation}
    n_{[\tau_a,\tau_b]} = \sum_{s,t=0}^\infty n_{st;[\tau_a,\tau_b]},
\end{equation}
where $n_{st;[\tau_a,\tau_b]}$ is defined as the number of pairs caused when Alice and Bob's states are $s$-photon states and $t$-photon states and the intensity choice is $[\tau_a,\tau_b]$. Define $p_{[\tau_a,\tau_b]}$ as the probability that a pair corresponds to the intensity choice $[\tau_a,\tau_b]$. These probabilities are constants and can be calculated as 
\begin{equation}
    p_{[\tau_a,\tau_b]} = \frac{1}{p_s} \sum_{\tau^a_j + \tau^a_k = \tau_a} \sum_{\tau^b_j + \tau^b_k= \tau_b} p_{\text{save}}^{(\tau^a_j,\tau^b_j)} p_{\text{save}}^{(\tau^a_k,\tau^b_k)} p_{\tau^a_j} p_{\tau^a_k} p_{\tau^b_j} p_{\tau^b_k},
\end{equation}
where $p_{\text{save}}^{(\tau^a_j,\tau^b_j)}$ and $p_{\text{save}}^{(\tau^a_k,\tau^b_k)}$ have been defined in Eq. (\ref{eq8}). Here, $p_s$ is the total probability which is defined as
\begin{equation}
    p_s = \sum_{[\tau_a,\tau_b]} \sum_{\tau^a_j + \tau^a_k = \tau_a} \sum_{\tau^b_j + \tau^b_k= \tau_b} p_{\text{save}}^{(\tau^a_j,\tau^b_j)} p_{\text{save}}^{(\tau^a_k,\tau^b_k)} p_{\tau^a_j} p_{\tau^a_k} p_{\tau^b_j} p_{\tau^b_k}.
\end{equation}
Hence we can obtain that $\sum_{[\tau_a,\tau_b]} p_{[\tau_a,\tau_b]}$ = 1. Actually, the value of $p_s$ is not required in the subsequent analysis, as only the ratios between different $p_{[\tau_a,\tau_b]}$ are used and consequently $p_s$ cancels out. The probability corresponding to $[\nu_a,\nu_b]$ can be divided into two cases, denoted as $p_{[\nu_a,\nu_b]^\prime}$ and $p_{[\nu_a,\nu_b]^{\prime\prime}}$, when Alice and Bob choose the vacuum states in the same round or different rounds, respectively. Hence these two probabilities can be shown as
\begin{equation}
    \begin{aligned}
        p_{[\nu_a,\nu_b]^\prime} &= \frac{2}{p_{s}} p_{\text{save}} p_{\nu_a} p_{o_a} p_{\nu_b} p_{o_b},\\
        p_{[\nu_a,\nu_b]^{\prime\prime}} &= \frac{2}{p_{s}} p_{\text{save}}^2 p_{\nu_a} p_{o_a} p_{\nu_b} p_{o_b}.\\
    \end{aligned}
\end{equation}

We define the linear combinations
\begin{equation}
    \begin{aligned}
        n_\mu \triangleq & \frac{n_{[\mu_a,\mu_b]}}{p_{0,\mu_a} p_{0,\mu_b} p_{[\mu_a,\mu_b]}} - \frac{n_{[o_a,\mu_b]}}{p_{0,o_a} p_{0,\mu_b} p_{[o_a,\mu_b]}} - \frac{n_{[\mu_a,o_b]}}{p_{0,\mu_a} p_{0,o_b} p_{[\mu_a,o_b]}} + \frac{n_{[o_a,o_b]}}{p_{0,o_a} p_{0,o_b} p_{[o_a,o_b]}},\\
    \end{aligned}
\end{equation}
and
\begin{equation}
    \begin{aligned}
        n_\nu \triangleq & \frac{n_{[\nu_a,\nu_b]^\prime}}{2p_{0,\nu_a} p_{0,\nu_b} p_{[\nu_a,\nu_b]^\prime}} + \frac{n_{[\nu_a,\nu_b]^{^{\prime\prime}}}}{2p_{0,\nu_a} p_{0,\nu_b} p_{[\nu_a,\nu_b]^{\prime\prime}}} - \frac{n_{[o_a,\nu_b]}}{p_{0,o_a} p_{0,\nu_b} p_{[o_a,\nu_b]}} \\
        & - \frac{n_{[\nu_a,o_b]}}{p_{0,\nu_a} p_{0,o_b} p_{[\nu_a,o_b]}} + \frac{n_{[o_a,o_b]}}{p_{0,o_a} p_{0,o_b} p_{[o_a,o_b]}}.
    \end{aligned}
\end{equation}
Then we can obtain the obtain the lower bound of $n_{11;[\mu_a,\mu_b]}$, which is just the lower bound of $\underline{n}_{11}^z$, as \cite{xie2023RN1097}
\begin{equation}
    \label{eq2}
    \underline{n}_{11}^z = \underline{n}_{11;[\mu_a,\mu_b]} = \frac{1}{\alpha_{11}} \bigg[\frac{p_{0,\nu_a} p_{0,\nu_b}}{p_{s_a,\nu_a} p_{s_b,\nu_b}} n_\nu - \frac{p_{0,\mu_a} p_{0,\mu_b}}{p_{s_a,\mu_a} p_{s_b,\mu_b}} n_\mu \bigg],
\end{equation}
where
\begin{equation}
    \alpha_{jk} = \frac{1}{p_{[\mu_a,\mu_b]}} \bigg[ \frac{p_{j,\nu_a} p_{k,\nu_b}}{p_{s_a,\nu_a} p_{s_b,\nu_b} p_{j,\mu_a} p_{k,\mu_b}} - \frac{1}{p_{s_a,\mu_a} p_{s_b,\mu_b}} \bigg],
\end{equation}
and
\begin{equation}
    \begin{cases}
        s_a=1,s_b=2, & \text{if } \nu_a \mu_b \leq \nu_b \mu_a, \\
        s_a=2,s_b=1, & \text{if } \nu_a \mu_b > \nu_b \mu_a.
    \end{cases}
\end{equation}

The number of error pairs $m_{[2\nu_a,2\nu_b]}$, which is defined in step (5) of Box 1, can be decomposed as
\begin{equation}
    m_{[2\nu_a,2\nu_b]} = \sum_{s,t=0}^\infty m_{st;[2\nu_a,2\nu_b]},
\end{equation}
where $m_{st;[2\nu_a,2\nu_b]}$ is defined as the number of error pairs caused when Alice and Bob's states are $s$-photon states and $t$-photon states and the intensity choice is $[2\nu_a,2\nu_b]$. We define the linear combination
\begin{equation}
    \label{eq3}
    \begin{aligned}
        m_{2\nu} \triangleq & \frac{m_{[2\nu_a,2\nu_b]}}{p_{0,2\nu_a} p_{0,2\nu_b} p_{[2\nu_a,2\nu_b]}} - \frac{n_{[o_a,2\nu_b]}}{2p_{0,o_a} p_{0,2\nu_b} p_{[o_a,2\nu_b]}} - \frac{n_{[2\nu_a,o_b]}}{2p_{0,2\nu_a} p_{0,o_b} p_{[2\nu_a,o_b]}} + \frac{n_{[o_a,o_b]}}{2p_{0,o_a} p_{0,o_b} p_{[o_a,o_b]}},
    \end{aligned}
\end{equation}
and can obtain the upper bound of $m_{11;[2\nu_a,2\nu_b]}$ as \cite{xie2023RN1097}
\begin{equation}
    \overline{m}_{11;[2\nu_a,2\nu_b]} \leq p_{0,2\nu_a} p_{0,2\nu_b} p_{[2\nu_a,2\nu_b]} m_{2\nu}.
\end{equation}
Since the lower bound of $n_{11;[2\nu_a,2\nu_b]}$ can be obtained based on Eq. (\ref{eq2}) as
\begin{equation}
    \underline{n}_{11;[2\nu_a,2\nu_b]} = \frac{p_{1,2\nu_a} p_{1,2\nu_b} p_{[2\nu_a,2\nu_b]}}{p_{1,\mu_a} p_{1,\mu_b} p_{[\mu_a,\mu_b]}} \underline{n}_{11;[\mu_a,\mu_b]},
\end{equation}
the upper bound of the phase error rate $e_{11}^x$ can be estimated as
\begin{equation}
    \overline{e}_{11}^x = \frac{\overline{m}_{1,1;[2\nu_a,2\nu_b]}}{\underline{n}_{1,1;[2\nu_a,2\nu_b]}}.
\end{equation}

In the finite case, the statistical fluctuation should be addressed. We could replace $n_\mu$, $n_\nu$ and $m_{2\nu}$ in Eqs. (\ref{eq2}) and (\ref{eq3}) with $n_\mu^\prime$, $n_\nu^\prime$ and $m_{2\nu}^\prime$ defined below to estimate the lower bound of $n_{11}^z$ and upper bound of $e_{11}^x$ \cite{xie2023RN1097}
\begin{equation}
    \begin{aligned}
        n_\mu^\prime \triangleq & \frac{\overline{n}^*_{[\mu_a,\mu_b]}}{p_{0,\mu_a} p_{0,\mu_b} p_{[\mu_a,\mu_b]}} - \frac{\underline{n}^*_{[o_a,\mu_b]}}{p_{0,o_a} p_{0,\mu_b} p_{[o_a,\mu_b]}} - \frac{\underline{n}^*_{[\mu_a,o_b]}}{p_{0,\mu_a} p_{0,o_b} p_{[\mu_a,o_b]}} + \frac{\underline{n}^*_{[o_a,o_b]}}{p_{0,o_a} p_{0,o_b} p_{[o_a,o_b]}},\\
        n_\nu^\prime \triangleq & \frac{p_{\text{save}} \underline{n}^*_{[\nu_a,\nu_b]^\prime} + \underline{n}^*_{[\nu_a,\nu_b]^{\prime\prime}}}{2 p_{0,\nu_a} p_{0,\nu_b} p_{[\nu_a,\nu_b]^{\prime\prime}}} - \frac{\overline{n}^*_{[o_a,\nu_b]}}{p_{0,o_a} p_{0,\nu_b} p_{[o_a,\nu_b]}} - \frac{\overline{n}^*_{[\nu_a,o_b]}}{p_{0,\nu_a} p_{0,o_b} p_{[\nu_a,o_b]}} + \frac{\underline{n}^*_{[o_a,o_b]}}{p_{0,o_a} p_{0,o_b} p_{[o_a,o_b]}},\\
    \end{aligned}
\end{equation}
and 
\begin{equation}
    \begin{aligned}
        m_{2\nu}^\prime \triangleq & \frac{m_{[2\nu_a,2\nu_b]}}{p_{0,2\nu_a} p_{0,2\nu_b} p_{[2\nu_a,2\nu_b]}} - \frac{\underline{n}^*_{[o_a,2\nu_b]}}{2p_{0,o_a} p_{0,2\nu_b} p_{[o_a,2\nu_b]}} - \frac{\underline{n}^*_{[2\nu_a,o_b]}}{2p_{0,2\nu_a} p_{0,o_b} p_{[2\nu_a,o_b]}} + \frac{\overline{n}^*_{[o_a,o_b]}}{2p_{0,o_a} p_{0,o_b} p_{[o_a,o_b]}}.
    \end{aligned}
\end{equation}
Here, the bounds of the expectation value $\underline{n}^*_{[\tau_a,\tau_b]}$ and $\overline{n}^*_{[\tau_a,\tau_b]}$ can be calculated with the observed value $n_{[\tau_a,\tau_b]}$ using the Chernoff bound \cite{zhang2017RN87}. Given an observed value $n$, we could obtain the lower and upper bound of the expectation value with success probability $1-\varepsilon_U$ and $1-\varepsilon_L$ as \cite{zhang2017RN87}
\begin{equation}
    \begin{aligned}
        \underline{n}^* &= \frac{n}{1 + \chi_L},\\
        \overline{n}^* &= \frac{n}{1 - \chi_U},
    \end{aligned}
\end{equation}
where $\chi_L$ and $\chi_U$ can be obtained by solving the following equations \cite{zhang2017RN87}
\begin{equation}
    \begin{aligned}
        & \bigg[ \frac{e^{\chi_L}}{(1 + \chi_L)^{1 + \chi_L}} \bigg]^{\frac{n}{1+\chi_L}} = \varepsilon_L,\\
        & \bigg[ \frac{e^{-\chi_U}}{(1 - \chi_U)^{1 - \chi_U}} \bigg]^{\frac{n}{1-\chi_U}} = \varepsilon_U.
    \end{aligned}
\end{equation}
Based on the random-sampling theory (without replacement) \cite{curty2014RN100,lim2014RN93}, we could guarantee the security in the finite case.

\section{Performance}
\label{performance}

To simulate the SKR, we set the parameters of SPDs as $\eta_{d_0} = \eta_{d_1} = 78\%$ and $p_{d_0} = p_{d_1} = 1$E$-8$, the intensities $\mu_a = \mu_b = 0.542$ and $\nu_a = \nu_b = 0.035$, and the probabilities $p_{\mu_a} = p_{\mu_b} = 0.261$ and $p_{\nu_a} = p_{\nu_b} = 0.344$ \cite{zhou2023RN978}. The quantum channels are set as the standard fiber with a fiber loss coefficient of 0.2 dB/km. And the fiber length between Alice and Charlie is set equal to that between Bob and Charlie. With these parameters, we simulate and compare the SKR with the original pairing strategy \cite{zeng2022RN816} and the flexible pairing strategy in this work. The detailed simulation method is present in App. \ref{simulation_method}.

\begin{figure}[t]
    \centering
    \includegraphics[width=0.6\textwidth]{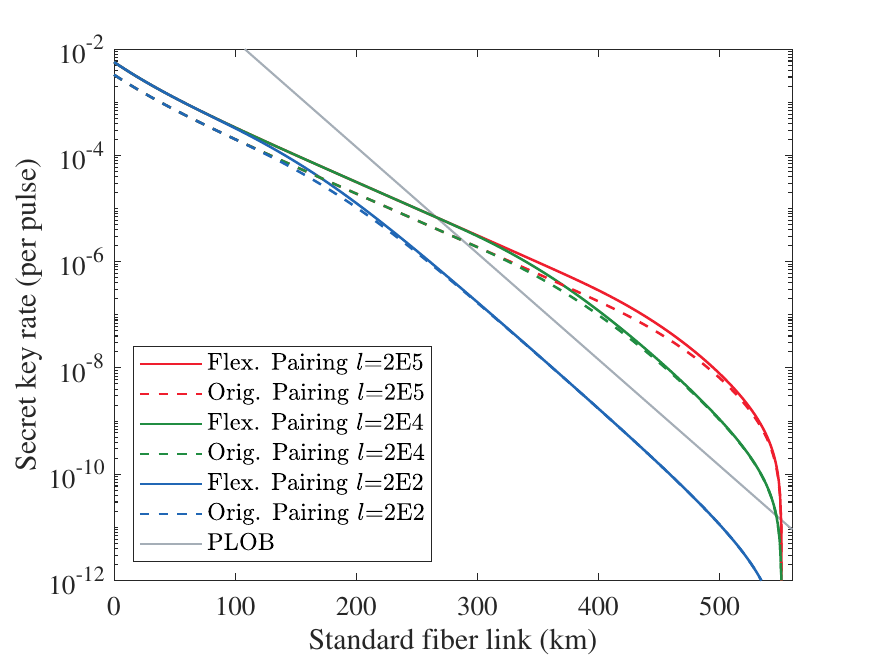}
    \caption{Secret key rate (per pulse) in logarithmic scale versus the transmission distance between Alice and Bob with different pairing intervals in the asymptotic case. The dashed lines are the results of the original pairing strategy in Ref. \cite{zeng2022RN816} and the solid lines are the results of the flexible pairing strategy in this work.}
    \label{fig_skr_asym}
\end{figure}

\begin{figure}[h]
    \centering
    \includegraphics[width=0.6\textwidth]{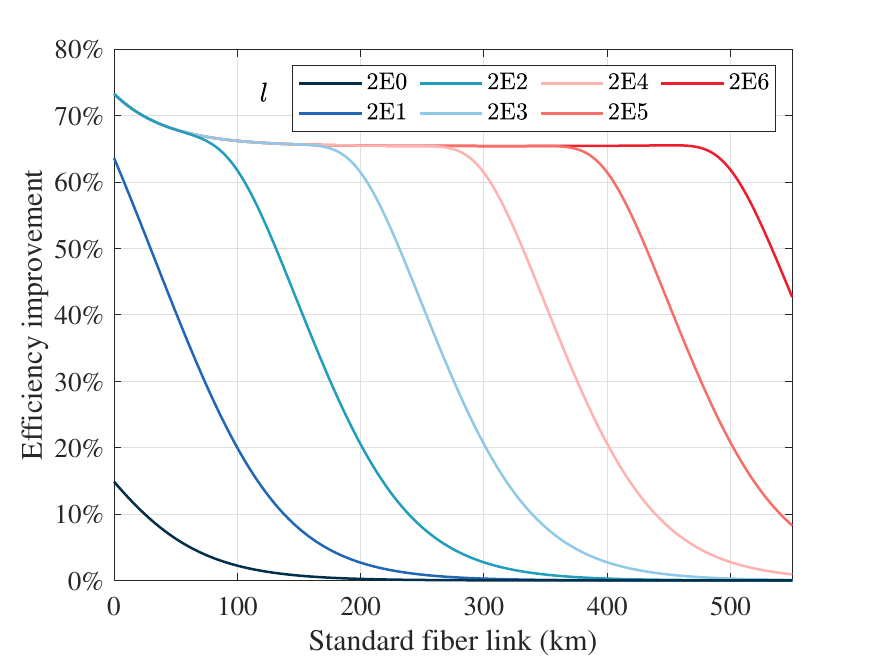}
    \caption{The improvement of the SKR with the flexible pairing strategy in this work compared with the original pairing strategy in Ref. \cite{zeng2022RN816}. Different lines are simulated with different pairing intervals in the asymptotic case.}
    \label{fig_ratio_asym}
\end{figure}

In Fig. \ref{fig_skr_asym}, we simulate the SKR with different pairing interval $l$ in the asymptotic case. The maximal pairing interval is set as 2E5, which is practical in the MP-QKD system \cite{zhou2023RN978}. Besides, we set the pairing interval to be more achievable as 2E4 and 2E2. The results show that the SKR is improved but may be limited at long distance when the pairing interval is small. With the reasonable pairing interval 2E5, the improvement is evident nearly below 500 km. For further study, we consider various pairing intervals and show the efficiency improvement of the SKR in Fig. \ref{fig_ratio_asym}. The overall trend shows that the improvement is high at close distance, approximately 65\% at medium distance, and gradually decreases at long distance. Furthermore, with an increase in the pairing interval, the distance at which the improvement starts to decrease also increases. The decrease in the improvement at long distance can be attributed to the insufficient number of effective rounds. Therefore, it is beneficial to address the issue of laser decoherence and phase drift in quantum channels to improve the SKR by increasing the pairing interval.

\begin{figure}[t]
    \centering
    \includegraphics[width=0.6\textwidth]{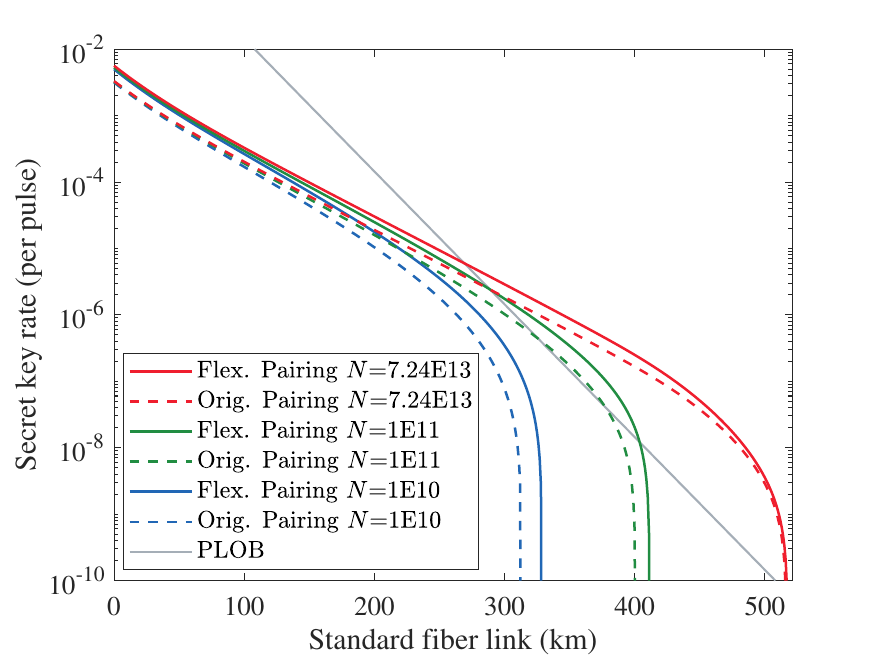}
    \caption{Secret key rate (per pulse) in logarithmic scale versus the transmission distance between Alice and Bob in the finite case. The dashed lines are the results of the original pairing strategy in Ref. \cite{zeng2022RN816} and the solid lines are the results of the flexible pairing strategy in this work.}
    \label{fig_skr_finite}
\end{figure}

\begin{figure}[t]
    \centering
    \includegraphics[width=0.6\textwidth]{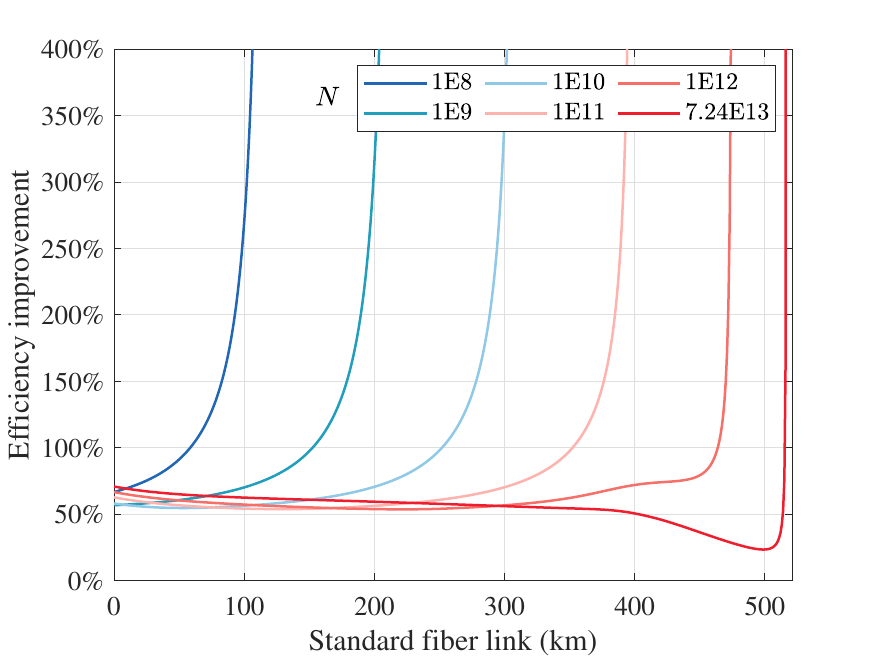}
    \caption{The improvement of the SKR with the flexible pairing strategy in this work compared with the original pairing strategy in Ref. \cite{zeng2022RN816}. Different lines are simulated with different block lengths $N$ in the finite case.}
    \label{fig_ratio_finite}
\end{figure}

In Fig. \ref{fig_skr_finite}, we compare the SKR with fixed pairing interval $l=2$E5 but different block lengths $N$ in the finite case. The maximal block length $N$ is set as 7.24E13 \cite{zhou2023RN978} and we also consider more achievable $N$ as 1E11 and 1E10. The results show that the improvement is more evident with smaller $N$. For example, not only the SKR is improved at the same distance but also the achievable distance is enhanced with $N=1$E11 and 1E10. The reason is that the achievable distance with small $N$ is limited due to the finite effects but not the insufficient effective rounds in a pairing interval. When the block length is large, the finite effects only become apparent at long distance. At this time, the effective rounds in a pairing interval are insufficient and hence the improvement is inapparent. In Fig. \ref{fig_ratio_finite}, we consider various block lengths $N$ and show the efficiency improvement of the SKR. The results show that the improvement is greater than 50\% and rises as the distance increases with block lengths 1E8, 1E9, 1E10, and 1E11. And the reason that the improvement is particularly high at long distance is that the SKR with the original pairing strategy reduces to 0. When considering large block lengths such as 1E12 and 7.24E13, the improvement shows a fine decrease at long distance and then increases rapidly at the end. As mentioned above, the decrease is owing to the insufficient effective rounds in a pairing interval. Overall, the SKR is greatly improved with the flexible pairing strategy.

\begin{figure}[t]
    \centering
    \includegraphics[width=0.6\textwidth]{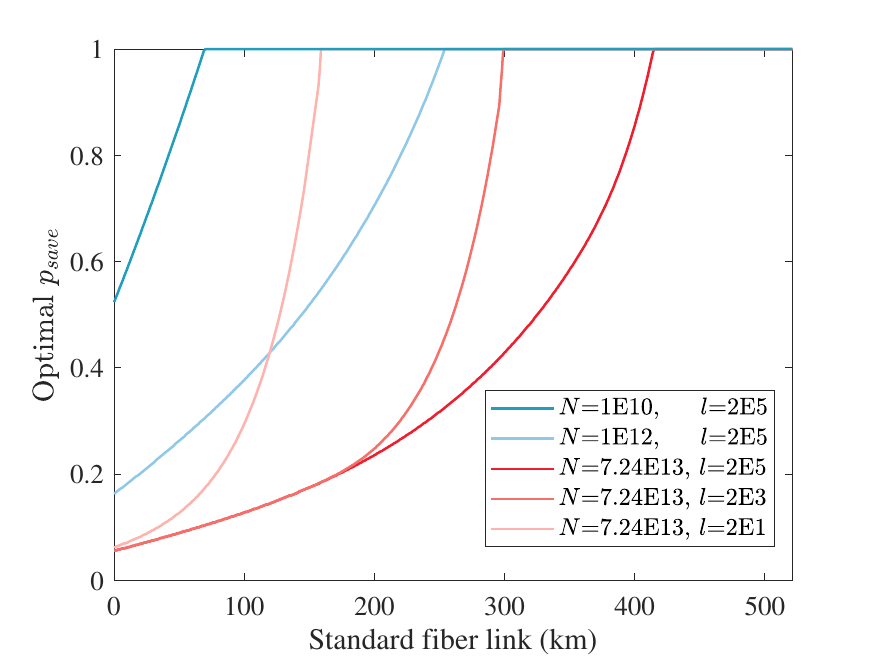}
    \caption{The optimal $p_{\text{save}}$ under various maximal block length $N$ and pairing interval $l$.}
    \label{fig_p_save}
\end{figure}

In the flexible pairing strategy in Box 2, the probability $p_{\text{save}}$ is important and should be optimized. In the asymptotic case, the optimal $p_{\text{save}} \rightarrow  0$ so as to guarantee high pairing efficiency of Z basis. However, the number of X basis is reduced with small $p_{\text{save}}$. Therefore, in the finite case, we should optimize $p_{\text{save}}$ to balance Z and X bases to obtain optimal SKR. In Fig. \ref{fig_p_save}, we show the optimal $p_{\text{save}}$ under various maximal block length $N$ and pairing interval $l$. The results show that as the channel loss increases, the optimal $p_{\text{save}}$ gradually increases towards 1. At a fixed $N$, a larger $l$ leads to a smaller optimal $p_{\text{save}}$. Conversely, at a fixed $l$, a larger $N$ also results in a smaller optimal $p_{\text{save}}$. This is because a larger $N$ or $l$ will result in a higher number of final pairs, allowing for a reduction in $p_{\text{save}}$ to achieve a higher key rate. When $N$ or $l$ decreases, to ensure the tightness of parameter estimation against statistical fluctuations, it is necessary to increase $p_{\text{save}}$ to guarantee the number of X basis pairs. Furthermore, as $N$ increases, the optimal $p_{\text{save}}$ at 0 km distance gradually decreases. In contrast, when $l$ increases, the optimal $p_{\text{save}}$ at 0 km remains nearly constant. Overall, the proposed method demonstrates more noticeable improvements under larger $N$ and $l$ conditions.

\begin{figure}[t]
    \centering
    \includegraphics[width=0.6\textwidth]{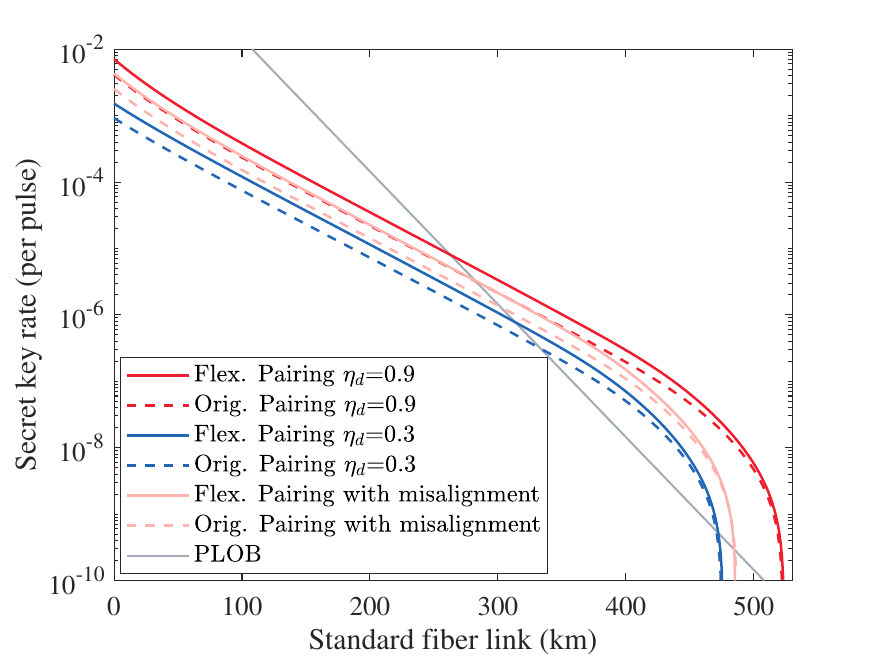}
    \caption{Secret key rate (per pulse) in logarithmic scale versus the transmission distance between Alice and Bob in various practical conditions.  The dashed lines are the results of the original pairing strategy in Ref. \cite{zeng2022RN816} and the solid lines are the results of the flexible pairing strategy in this work. The red and blue lines are obtained by setting $\eta_{d_0} = \eta_{d_1} = \eta$ without misalignment. The pink lines are obtained by setting the interference misalignment error rate $E_{\text{HOM}} = 0.04$, the laser frequency difference $\Delta f = 10$ Hz, and the fiber phase drift rate $\omega_{\text{fiber}} = $5.9E3 rad/s.}
    \label{fig_factor}
\end{figure}

\begin{figure}[t]
    \centering
    \includegraphics[width=0.6\textwidth]{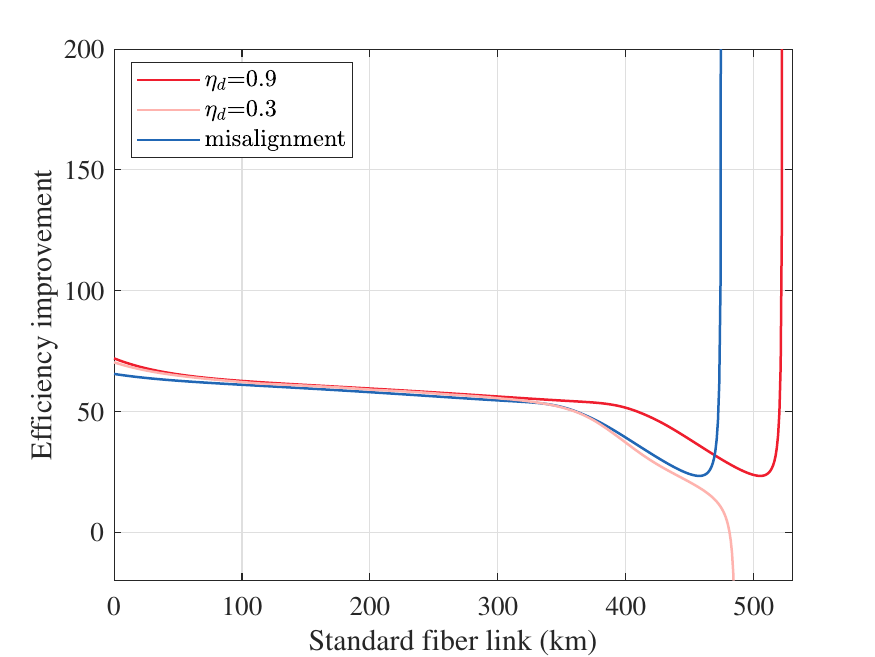}
    \caption{The improvement of the SKR with the flexible pairing strategy in this work compared with the original pairing strategy in Ref. \cite{zeng2022RN816} in various practical conditions.}
    \label{fig_factor_ratio}
\end{figure}

Finally, we analyze the proposed pairing strategy under various practical conditions. The number of pairs and the error rate that directly determine the SKR of MP-QKD, which are affected by many factors including the detection efficiency, the pairing interval, the laser coherence stability, the interference contrast, and the phase drift in the channel, etc. The detection efficiency and the pairing interval will affect the pairing efficiency, while other factors determine the error rate. In Fig. \ref{fig_factor}, we present the comparative results by changing the simulation settings. As the pairing interval has been analyzed in Fig. \ref{fig_skr_asym}, we first only consider different detection efficiencies, such as a high value of 0.9 or a low value of 0.3 in Fig. \ref{fig_factor}. Besides, we consider the misalignment, including the interference misalignment error rate $E_{\text{HOM}}$, the laser frequency difference $\Delta f$, and the fiber phase drift rate $\omega_{\text{fiber}}$. We set the system clock frequency $F$ as 1E9. The parameters of the misalignment are set according to Ref. \cite{zhou2023RN978}. The results show that the proposed pairing strategy is robust and capable of improving the SKR under various practical conditions. In Fig. \ref{fig_factor_ratio}, we present the efficiency improvement of the SKR. It shows that the basic improvement is greater than 55\%. At long distance with different detection efficiencies, the improvement first decreases and then rapidly increases. When considering the misalignment, the improvement in the key rate may decrease and even become lower than the original pairing strategy. This is because the flexible pairing strategy discards some invalid rounds, leading to an increase in the average pairing interval, which in turn causes a slight increase in the error rate. However, this decrease occurs only in the final few kilometers. Overall, the proposed method still significantly enhances the protocol's performance.

Based on the above simulation results, we can see that our method achieves significant improvements under certain conditions, while its effectiveness may be limited under others. Therefore, it is beneficial to analyze the practical requirements and limitations so as to achieve better performance. In short and medium distance experiments (e.g., within 300 km), the block length and the pairing interval do not need to be excessively large. According to Fig. \ref{fig_skr_finite}, the performance improvement under these conditions is particularly significant. In contrast, under long distance conditions, due to the limited amount of effective rounds, a larger pairing interval is required to achieve significant improvement, as shown in Fig. \ref{fig_skr_asym}. However, according to the results in Fig. \ref{fig_factor}, when practical factors are considered, an excessively large pairing interval may lead to an increase in the error rate, thereby limiting the overall performance enhancement. Therefore, it is important to evaluate and surpass the system's noise level to obtain optimal performance. To surpass the system's noise, it is crucial to ensure laser wavelength consistency and the stability of quantum channel, thereby suppressing the error rate.

\section{Conclusion}
\label{conclusion}

MP-QKD is a promising protocol that enjoys both simple implementation and high performance. The post-measurement pairing plays a critical role in both security and performance. In this work, we present an entanglement model for the decoy-state MP-QKD. With this model, we design a flexible pairing strategy for decoy-state MP-QKD and rigorously prove its security. The simulation results demonstrate the enhancement of SKR in both asymptotic and finite cases. The achievable distance can be extended in the finite case, especially with a small block length. This promotes the practicality of decoy-state MP-QKD for long-distance secure communication. Besides, the entanglement model provides a theoretical framework for further security and performance analysis of decoy-state MP-QKD. For future work, the proposed method could be combined with existing approaches, such as the optimization of the pairing interval \cite{zhou2025RN1272}, improving the pairing efficiency of the X basis \cite{zhou2025RN1275}, and analysis under practical conditions \cite{wang2023RN1051,liu2023RN1182,lu2024RN1183,li2024RN1271,zhou2024RN1084}, in order to further enhance the performance of the MP-QKD.

\section*{Funding}
National Key Research and Development Program of China (No. 2020YFA0309702), Henan Science and Technology Major Project of the Department of Science \& Technology of Henan Province (No. 241100210400), National Natural Science Foundation of China (Nos. U2130205, 62101597), Natural Science Foundation of Henan (No. 242300421219).

\appendix

\section{Simulation method}
\label{simulation_method}

In this section, we give the method to simulate the performance of MP-QKD with the pairing strategy in Box 2. The key is to simulate the number of pairs $n_{[\tau_a,\tau_b]}$ and $m_{[\tau_a,\tau_b]}$. With these parameters, we could estimate $\underline{n}_{11}^z$ and $\overline{e}_{11}^x$ with the estimation method in Sec. \ref{sec_parameter_estimation}, and calculate the SKR in Eq. (\ref{eq4}).

We first analyze the situation in a single round. Suppose Alice and Bob's intensities are $\tau_k^a$ and $\tau_k^b$ and the phase difference is $\delta$ in the $k$-th round, the effective intensities after transmission and interference at two SPDs can be simulated as
\begin{equation}
    \begin{aligned}
        \tau_L & = \frac{\eta_{d_0}}{2} \Big(\eta_a\tau_k^a + \eta_b\tau_k^b + 2\sqrt{\eta_a\tau_k^a\eta_b\tau_k^b} \cos\delta \Big),\\
        \tau_R & = \frac{\eta_{d_1}}{2} \Big(\eta_a\tau_k^a + \eta_b\tau_k^b - 2\sqrt{\eta_a\tau_k^a\eta_b\tau_k^b} \cos\delta \Big),
    \end{aligned}
\end{equation}
where $\eta_{d_0}$ and $\eta_{d_1}$ are the detection efficiencies of the SPDs, $\eta_a$ ($\eta_b$) is the transmission efficiency of the quantum channel between Alice (Bob) and Charlie. Then the probabilities $q_{\tau_k^a,\tau_k^b}^{\delta,L}$ or $q_{\tau_k^a,\tau_k^b}^{\delta,R}$ that only the left or right SPD clicks can be expressed as
\begin{equation}
    \begin{aligned}
        q_{\tau_k^a,\tau_k^b}^{\delta,L} &= [1 - (1 - p_{d_0}) e^{-\tau_L}] (1 - p_{d_1}) e^{-\tau_R},\\
        q_{\tau_k^a,\tau_k^b}^{\delta,R} &= (1 - p_{d_0}) e^{-\tau_L} [1 - (1 - p_{d_1}) e^{-\tau_R}],
    \end{aligned}
\end{equation}
where $p_{d_0}$ and $p_{d_1}$ are the dark count probabilities of the SPDs. The average probability $q_{\tau_k^a,\tau_k^b}$ that a round is effective regarding the intensities $\tau_k^a,\tau_k^b$ can be calculated as
\begin{equation}
    q_{\tau_k^a,\tau_k^b} = \frac{1}{2\pi} \int_{0}^{2\pi} \Big( q_{\tau_k^a,\tau_k^b}^{\delta,L} + q_{\tau_k^a,\tau_k^b}^{\delta,R} \Big) d\delta.
\end{equation}
Therefore, the total average probability $q_0$ that a round is effective can be given as
\begin{equation}
    q_0 = \sum_{\tau_k^a,\tau_k^b} p_{\tau_k^a} p_{\tau_k^b} q_{\tau_k^a,\tau_k^b}.
\end{equation}
In Box 2, the pairing strategy will first assign a temporary parameter to filter out partial rounds. The collective probability $q$ that a round is effective and saved after filtering can be shown as
\begin{equation}
    q = \sum_{\tau_k^a,\tau_k^b} p_{\tau_k^a} p_{\tau_k^b} q_{\tau_k^a,\tau_k^b} p_{\text{save}}^{(\tau_k^a,\tau_k^b)}.
\end{equation}

To analyze the number of pairs, we analyze the pairing efficiency $r$, which is defined as the ratio of the pairs among all rounds $N$. Suppose the pairing interval is $l$, the pairing efficiency of all pairs can be given as \cite{zeng2022RN816}
\begin{equation}
    r = \bigg[ \frac{1}{q} + \frac{1}{q} \frac{1}{1 - (1 - q)^l} \bigg]^{-1}.
\end{equation}
Hence the total number of pairs is $n_{\text{tot}} = N r$. The number of different pairs can be calculated as
\begin{equation}
    n_{[\tau_a,\tau_b]} = \frac{n_{\text{tot}}}{q^2} \sum_{\tau_j^a + \tau_k^a = \tau_a} \sum_{\tau_j^b + \tau_k^b = \tau_b} p_{\tau^a_j} p_{\tau^b_j} p_{\tau^a_k} p_{\tau^b_k} q_{\tau^a_j, \tau^b_j} q_{\tau^a_k, \tau^b_k} p_{\text{save}}^{(\tau_j^a,\tau_j^b)} p_{\text{save}}^{(\tau_k^a,\tau_k^b)}.
\end{equation}
The special cases $n_{[\nu_a,\nu_b]^{\prime}}$ and $n_{[\nu_a,\nu_b]^{\prime\prime}}$ can be calculated as
\begin{equation}
    \begin{aligned}
        n_{[\nu_a,\nu_b]^{\prime}} = \frac{n_{\text{tot}}}{q^2} 2 p_{\nu_a} p_{o_a} p_{\nu_b} p_{o_b} q_{\nu_a,\nu_b} q_{o_a,o_b} p_{\text{save}},\\
        n_{[\nu_a,\nu_b]^{\prime\prime}} = \frac{n_{\text{tot}}}{q^2} 2 p_{\nu_a} p_{o_a} p_{\nu_b} p_{o_b} q_{\nu_a,o_b} q_{o_a,\nu_b} p_{\text{save}}^2.
    \end{aligned}
\end{equation}
The number of error pairs when both Alice and Bob are $Z$ basis can be calculated as
\begin{equation}
    m_{[\mu_a,\mu_b]}= 2 \frac{n_{\text{tot}}}{q^2} p_{\mu_a} p_{\mu_b} p_{o_a} p_{o_b} q_{\mu_a, \mu_b} q_{o_b, o_a}.
\end{equation}
As the pairs in $X$ basis are sifted according to the phase differences, the number of pairs in $X$ basis is shown as
\begin{equation}
    \begin{aligned}
        n_{[2\nu_a,2\nu_b]} &= \frac{n_{\text{tot}}}{q^2} \frac{2}{M} \frac{1}{2\pi} p_{\nu_a}^2 p_{\nu_b}^2 p_{\text{save}}^2 \int_{0}^{2\pi} \Big[ q_{\nu_a, \nu_b}^{\delta,L} + q_{\nu_a, \nu_b}^{\delta,R} \Big]^2 d\delta,\\
        m_{[2\nu_a,2\nu_b]} &= 2 \frac{n_{\text{tot}}}{q^2} \frac{2}{M} \frac{1}{2\pi} p_{\nu_a}^2 p_{\nu_b}^2 p_{\text{save}}^2 \int_{0}^{2\pi} q_{\nu_a, \nu_b}^{\delta,L} q_{\nu_a, \nu_b}^{\delta,R} d\delta,
    \end{aligned}
\end{equation}
where the coefficient $2/M$ is due to the sifting of the phase slices in step (5). The bit error rate in Z basis is $E_z = m_{[\mu_a,\mu_b]} / n_{[\mu_a,\mu_b]}$ and the amount of information revealed in the error correction step is $\lambda_{\text{EC}} = n_{[\mu_a,\mu_b]} f H_2(E_z)$, where $f$ is the efficiency of the error correction.

In addition, when we consider the misalignment, the number of error pairs in X basis is re-expressed as \cite{zhou2023RN978}
\begin{equation}
    \begin{aligned}
        m_{[2\nu_a,2\nu_b]} = & \frac{n_{\text{tot}}}{q^2} \frac{2}{M} \frac{1}{2\pi} p_{\nu_a}^2 p_{\nu_b}^2 p_{\text{save}}^2 \int_{0}^{2\pi} \Big[ (1 - E_{\text{HOM}}) \big(q_{\nu_a, \nu_b}^{\delta,L} q_{\nu_a, \nu_b}^{\delta+\vartheta,R} + q_{\nu_a, \nu_b}^{\delta,R} q_{\nu_a, \nu_b}^{\delta+\vartheta,L} \big) \\
        &+ E_{\text{HOM}} \big(q_{\nu_a, \nu_b}^{\delta,L} q_{\nu_a, \nu_b}^{\delta+\vartheta,L} + q_{\nu_a, \nu_b}^{\delta,R} q_{\nu_a, \nu_b}^{\delta+\vartheta,R}\big) \Big] d\delta ,
    \end{aligned}
\end{equation}
where $E_{\text{HOM}}$ is the interference misalignment error rate, $\vartheta = T_\text{mean} (2\pi \Delta f + \omega_{\text{fiber}})$ is the phase misalignment with the average of paring interval $T_{\text{mean}}$, the laser frequency difference $\Delta f$ and the phase drift rate in fiber $\omega_{\text{fiber}}$. Here, the average of the paring interval can be simulated as \cite{zhou2023RN978}
\begin{equation}
    T_{\text{mean}} = \frac{1 - l q \{[1 - (1 - q)^{l}]^{-1} - 1\}}{qF}.
\end{equation}

\bibliographystyle{spphys}
\bibliography{MP_Pairing_Ref}

\begin{thebibliography}{10}
\providecommand{\url}[1]{{#1}}
\providecommand{\urlprefix}{URL }
\expandafter\ifx\csname urlstyle\endcsname\relax
  \providecommand{\doi}[1]{DOI \discretionary{}{}{}#1}\else
  \providecommand{\doi}{DOI \discretionary{}{}{}\begingroup \urlstyle{rm}\Url}\fi

\bibitem{bennett2014RN153}
C.H. Bennett, G.~Brassard, Theoretical Computer Science \textbf{560}, 7 (2014)

\bibitem{xu2020RN130}
F.~Xu, X.~Ma, Q.~Zhang, H.K. Lo, J.W. Pan, Reviews of Modern Physics \textbf{92}(2), 025002 (2020)

\bibitem{pirandola2020RN485}
S.~Pirandola, U.L. Andersen, L.~Banchi, M.~Berta, D.~Bunandar, R.~Colbeck, D.~Englund, T.~Gehring, C.~Lupo, C.~Ottaviani, J.L. Pereira, M.~Razavi, J.~Shamsul~Shaari, M.~Tomamichel, V.C. Usenko, G.~Vallone, P.~Villoresi, P.~Wallden, Advances in Optics and Photonics \textbf{12}(4), 1012 (2020)

\bibitem{yin2016RN288}
H.L. Yin, T.Y. Chen, Z.W. Yu, H.~Liu, L.X. You, Y.H. Zhou, S.J. Chen, Y.~Mao, M.Q. Huang, W.J. Zhang, H.~Chen, M.J. Li, D.~Nolan, F.~Zhou, X.~Jiang, Z.~Wang, Q.~Zhang, X.B. Wang, J.W. Pan, Physical Review Letters \textbf{117}(19), 190501 (2016)

\bibitem{boaron2018RN286}
A.~Boaron, G.~Boso, D.~Rusca, C.~Vulliez, C.~Autebert, M.~Caloz, M.~Perrenoud, G.~Gras, F.~Bussières, M.J. Li, D.~Nolan, A.~Martin, H.~Zbinden, Physical Review Letters \textbf{121}(19), 190502 (2018)

\bibitem{liao2018RN512}
S.K. Liao, W.Q. Cai, J.~Handsteiner, B.~Liu, J.~Yin, L.~Zhang, D.~Rauch, M.~Fink, J.G. Ren, W.Y. Liu, Y.~Li, Q.~Shen, Y.~Cao, F.Z. Li, J.F. Wang, Y.M. Huang, L.~Deng, T.~Xi, L.~Ma, T.~Hu, L.~Li, N.L. Liu, F.~Koidl, P.~Wang, Y.A. Chen, X.B. Wang, M.~Steindorfer, G.~Kirchner, C.Y. Lu, R.~Shu, R.~Ursin, T.~Scheidl, C.Z. Peng, J.Y. Wang, A.~Zeilinger, J.W. Pan, Physical Review Letters \textbf{120}(3), 030501 (2018)

\bibitem{pittaluga2021RN482}
M.~Pittaluga, M.~Minder, M.~Lucamarini, M.~Sanzaro, R.I. Woodward, M.J. Li, Z.~Yuan, A.J. Shields, Nature Photonics \textbf{15}(7), 530 (2021)

\bibitem{wang2022RN585}
S.~Wang, Z.Q. Yin, D.Y. He, W.~Chen, R.Q. Wang, P.~Ye, Y.~Zhou, G.J. Fan-Yuan, F.X. Wang, W.~Chen, Y.G. Zhu, P.V. Morozov, A.V. Divochiy, Z.~Zhou, G.C. Guo, Z.F. Han, Nature Photonics \textbf{16}(2), 154 (2022)

\bibitem{chen2022RN735}
J.P. Chen, C.~Zhang, Y.~Liu, C.~Jiang, D.F. Zhao, W.J. Zhang, F.X. Chen, H.~Li, L.X. You, Z.~Wang, Y.~Chen, X.B. Wang, Q.~Zhang, J.W. Pan, Physical Review Letters \textbf{128}(18), 180502 (2022)

\bibitem{liu2023RN975}
Y.~Liu, W.J. Zhang, C.~Jiang, J.P. Chen, C.~Zhang, W.X. Pan, D.~Ma, H.~Dong, J.M. Xiong, C.J. Zhang, H.~Li, R.C. Wang, J.~Wu, T.Y. Chen, L.~You, X.B. Wang, Q.~Zhang, J.W. Pan, Physical Review Letters \textbf{130}(21), 210801 (2023)

\bibitem{yuan2018RN513}
Z.~Yuan, A.~Plews, R.~Takahashi, K.~Doi, W.~Tam, A.W. Sharpe, A.R. Dixon, E.~Lavelle, J.F. Dynes, A.~Murakami, M.~Kujiraoka, M.~Lucamarini, Y.~Tanizawa, H.~Sato, A.J. Shields, Journal of Lightwave Technology \textbf{36}(16), 3427 (2018)

\bibitem{fadri2023RN1023}
F.~Grünenfelder, A.~Boaron, G.V. Resta, M.~Perrenoud, D.~Rusca, C.~Barreiro, R.~Houlmann, R.~Sax, L.~Stasi, S.~El-Khoury, E.~Hänggi, N.~Bosshard, F.~Bussières, H.~Zbinden, Nature Photonics \textbf{17}(5), 422 (2023)

\bibitem{li2023RN987}
W.~Li, L.~Zhang, H.~Tan, Y.~Lu, S.K. Liao, J.~Huang, H.~Li, Z.~Wang, H.K. Mao, B.~Yan, Q.~Li, Y.~Liu, Q.~Zhang, C.Z. Peng, L.~You, F.~Xu, J.W. Pan, Nature Photonics \textbf{17}(5), 416 (2023)

\bibitem{ma2016RN1170}
C.~Ma, W.D. Sacher, Z.~Tang, J.C. Mikkelsen, Y.~Yang, F.~Xu, T.~Thiessen, H.K. Lo, J.K.S. Poon, Optica \textbf{3}(11), 1274 (2016)

\bibitem{sibson2017RN1190}
P.~Sibson, C.~Erven, M.~Godfrey, S.~Miki, T.~Yamashita, M.~Fujiwara, M.~Sasaki, H.~Terai, M.G. Tanner, C.M. Natarajan, R.H. Hadfield, J.L. O’Brien, M.G. Thompson, Nature Communications \textbf{8}(1), 13984 (2017)

\bibitem{bunandar2018RN1195}
D.~Bunandar, A.~Lentine, C.~Lee, H.~Cai, C.M. Long, N.~Boynton, N.~Martinez, C.~DeRose, C.~Chen, M.~Grein, D.~Trotter, A.~Starbuck, A.~Pomerene, S.~Hamilton, F.N.C. Wong, R.~Camacho, P.~Davids, J.~Urayama, D.~Englund, Physical Review X \textbf{8}(2), 021009 (2018)

\bibitem{zhang2019RN1188}
G.~Zhang, J.Y. Haw, H.~Cai, F.~Xu, S.M. Assad, J.F. Fitzsimons, X.~Zhou, Y.~Zhang, S.~Yu, J.~Wu, W.~Ser, L.C. Kwek, A.Q. Liu, Nature Photonics \textbf{13}(12), 839 (2019)

\bibitem{wei2020RN1030}
K.~Wei, W.~Li, H.~Tan, Y.~Li, H.~Min, W.J. Zhang, H.~Li, L.~You, Z.~Wang, X.~Jiang, T.Y. Chen, S.K. Liao, C.Z. Peng, F.~Xu, J.W. Pan, Physical Review X \textbf{10}(3), 031030 (2020)

\bibitem{semenenko2020RN1171}
H.~Semenenko, P.~Sibson, A.~Hart, M.G. Thompson, J.G. Rarity, C.~Erven, Optica \textbf{7}(3), 238 (2020)

\bibitem{wang2020RN1135}
F.X. Wang, W.~Wang, R.~Niu, X.~Wang, C.L. Zou, C.H. Dong, B.E. Little, S.T. Chu, H.~Liu, P.~Hao, S.~Liu, S.~Wang, Z.Q. Yin, D.Y. He, W.~Zhang, W.~Zhao, Z.F. Han, G.C. Guo, W.~Chen, Laser and Photonics Reviews \textbf{14}(2), 1900190 (2020)

\bibitem{elliott2005RN1087}
C.~Elliott, A.~Colvin, D.~Pearson, O.~Pikalo, J.~Schlafer, H.~Yeh, Proceedings of SPIE - The International Society for Optical Engineering  (2005)

\bibitem{stucki2011RN1086}
D.~Stucki, M.~Legré, F.~Buntschu, B.~Clausen, N.~Felber, N.~Gisin, L.~Henzen, P.~Junod, G.~Litzistorf, P.~Monbaron, L.~Monat, J.B. Page, D.~Perroud, G.~Ribordy, A.~Rochas, S.~Robyr, J.~Tavares, R.~Thew, P.~Trinkler, S.~Ventura, R.~Voirol, N.~Walenta, H.~Zbinden, New Journal of Physics \textbf{13}(12), 123001 (2011)

\bibitem{sasaki2011RN1085}
M.~Sasaki, M.~Fujiwara, H.~Ishizuka, W.~Klaus, K.~Wakui, M.~Takeoka, S.~Miki, T.~Yamashita, Z.~Wang, A.~Tanaka, K.~Yoshino, Y.~Nambu, S.~Takahashi, A.~Tajima, A.~Tomita, T.~Domeki, T.~Hasegawa, Y.~Sakai, H.~Kobayashi, T.~Asai, K.~Shimizu, T.~Tokura, T.~Tsurumaru, M.~Matsui, T.~Honjo, K.~Tamaki, H.~Takesue, Y.~Tokura, J.F. Dynes, A.R. Dixon, A.W. Sharpe, Z.L. Yuan, A.J. Shields, S.~Uchikoga, M.~Legré, S.~Robyr, P.~Trinkler, L.~Monat, J.B. Page, G.~Ribordy, A.~Poppe, A.~Allacher, O.~Maurhart, T.~Länger, M.~Peev, A.~Zeilinger, Optics Express \textbf{19}(11), 10387 (2011)

\bibitem{wang2014RN1217}
S.~Wang, W.~Chen, Z.Q. Yin, H.W. Li, D.Y. He, Y.H. Li, Z.~Zhou, X.T. Song, F.Y. Li, D.~Wang, H.~Chen, Y.G. Han, J.Z. Huang, J.F. Guo, P.L. Hao, M.~Li, C.M. Zhang, D.~Liu, W.Y. Liang, C.H. Miao, P.~Wu, G.C. Guo, Z.F. Han, Optics Express \textbf{22}(18), 21739 (2014)

\bibitem{chen2021RN519}
Y.A. Chen, Q.~Zhang, T.Y. Chen, W.Q. Cai, S.K. Liao, J.~Zhang, K.~Chen, J.~Yin, J.G. Ren, Z.~Chen, S.L. Han, Q.~Yu, K.~Liang, F.~Zhou, X.~Yuan, M.S. Zhao, T.Y. Wang, X.~Jiang, L.~Zhang, W.Y. Liu, Y.~Li, Q.~Shen, Y.~Cao, C.Y. Lu, R.~Shu, J.Y. Wang, L.~Li, N.L. Liu, F.~Xu, X.B. Wang, C.Z. Peng, J.W. Pan, Nature \textbf{589}(7841), 214 (2021)

\bibitem{pirandola2009RN1300}
S.~Pirandola, R.~García-Patrón, S.L. Braunstein, S.~Lloyd, Physical Review Letters \textbf{102}(5), 050503 (2009)

\bibitem{takeoka2014RN231}
M.~Takeoka, S.~Guha, M.M. Wilde, Nature Communications \textbf{5}(1), 5235 (2014)

\bibitem{pirandola2017RN103}
S.~Pirandola, R.~Laurenza, C.~Ottaviani, L.~Banchi, Nature Communications \textbf{8}(1), 15043 (2017)

\bibitem{das2021RN719}
S.~Das, S.~Bäuml, M.~Winczewski, K.~Horodecki, Physical Review X \textbf{11}(4), 041016 (2021)

\bibitem{lucamarini2018RN45}
M.~Lucamarini, Z.L. Yuan, J.F. Dynes, A.J. Shields, Nature \textbf{557}(7705), 400 (2018)

\bibitem{ma2018RN56}
X.~Ma, P.~Zeng, H.~Zhou, Physical Review X \textbf{8}(3), 031043 (2018)

\bibitem{wang2018RN22}
X.B. Wang, Z.W. Yu, X.L. Hu, Physical Review A \textbf{98}(6), 062323 (2018)

\bibitem{curty2019RN57}
M.~Curty, K.~Azuma, H.K. Lo, npj Quantum Information \textbf{5}(1), 64 (2019)

\bibitem{maeda2019RN507}
K.~Maeda, T.~Sasaki, M.~Koashi, Nature Communications \textbf{10}(1), 3140 (2019)

\bibitem{cui2019RN41}
C.~Cui, Z.Q. Yin, R.~Wang, W.~Chen, S.~Wang, G.C. Guo, Z.F. Han, Physical Review Applied \textbf{11}(3), 034053 (2019)

\bibitem{zhong2019RN52}
X.~Zhong, J.~Hu, M.~Curty, L.~Qian, H.K. Lo, Physical Review Letters \textbf{123}(10), 100506 (2019)

\bibitem{fang2020RN55}
X.T. Fang, P.~Zeng, H.~Liu, M.~Zou, W.~Wu, Y.L. Tang, Y.J. Sheng, Y.~Xiang, W.~Zhang, H.~Li, Z.~Wang, L.~You, M.J. Li, H.~Chen, Y.A. Chen, Q.~Zhang, C.Z. Peng, X.~Ma, T.Y. Chen, J.W. Pan, Nature Photonics \textbf{14}(7), 422 (2020)

\bibitem{chen2021RN586}
J.P. Chen, C.~Zhang, Y.~Liu, C.~Jiang, W.J. Zhang, Z.Y. Han, S.Z. Ma, X.L. Hu, Y.H. Li, H.~Liu, F.~Zhou, H.F. Jiang, T.Y. Chen, H.~Li, L.X. You, Z.~Wang, X.B. Wang, Q.~Zhang, J.W. Pan, Nature Photonics \textbf{15}(8), 570 (2021)

\bibitem{li2023RN979}
W.~Li, L.~Zhang, Y.~Lu, Z.P. Li, C.~Jiang, Y.~Liu, J.~Huang, H.~Li, Z.~Wang, X.B. Wang, Q.~Zhang, L.~You, F.~Xu, J.W. Pan, Physical Review Letters \textbf{130}(25), 250802 (2023)

\bibitem{zhou2023RN949}
L.~Zhou, J.~Lin, Y.~Jing, Z.~Yuan, Nature Communications \textbf{14}(1), 928 (2023)

\bibitem{chen2024RN1108}
J.P. Chen, F.~Zhou, C.~Zhang, C.~Jiang, F.X. Chen, J.~Huang, H.~Li, L.X. You, X.B. Wang, Y.~Liu, Q.~Zhang, J.W. Pan, Physical Review Letters \textbf{132}(26), 260802 (2024)

\bibitem{zhou2024RN1175}
L.~Zhou, J.~Lin, C.~Ge, Y.~Fan, Z.~Yuan, H.~Dong, Y.~Liu, D.~Ma, J.P. Chen, C.~Jiang, X.B. Wang, L.X. You, Q.~Zhang, J.W. Pan, Physical Review Applied \textbf{22}(6), 064057 (2024)

\bibitem{zeng2022RN816}
P.~Zeng, H.~Zhou, W.~Wu, X.~Ma, Nature Communications \textbf{13}(1), 3903 (2022)

\bibitem{xie2022RN724}
Y.M. Xie, Y.S. Lu, C.X. Weng, X.Y. Cao, Z.Y. Jia, Y.~Bao, Y.~Wang, Y.~Fu, H.L. Yin, Z.B. Chen, PRX Quantum \textbf{3}(2), 020315 (2022)

\bibitem{lo2012RN72}
H.K. Lo, M.~Curty, B.~Qi, Physical Review Letters \textbf{108}(13), 130503 (2012)

\bibitem{braunstein2012RN581}
S.L. Braunstein, S.~Pirandola, Physical Review Letters \textbf{108}(13), 130502 (2012)

\bibitem{ma2012RN73}
X.~Ma, M.~Razavi, Physical Review A \textbf{86}(6), 062319 (2012)

\bibitem{lu2025RN1293}
Y.F. Lu, Y.~Wang, H.W. Li, M.S. Jiang, X.X. Zhang, Y.Y. Zhang, Y.~Zhou, X.L. Jiang, H.T. Wang, Y.M. Zhao, C.~Zhou, W.S. Bao, Physical Review Research \textbf{7}(2), 023102 (2025)

\bibitem{zhu2023RN928}
H.T. Zhu, Y.~Huang, H.~Liu, P.~Zeng, M.~Zou, Y.~Dai, S.~Tang, H.~Li, L.~You, Z.~Wang, Y.A. Chen, X.~Ma, T.Y. Chen, J.W. Pan, Physical Review Letters \textbf{130}(3), 030801 (2023)

\bibitem{zhou2023RN978}
L.~Zhou, J.~Lin, Y.M. Xie, Y.S. Lu, Y.~Jing, H.L. Yin, Z.~Yuan, Physical Review Letters \textbf{130}(25), 250801 (2023)

\bibitem{zhu2024RN1180}
H.T. Zhu, Y.~Huang, W.X. Pan, C.W. Zhou, J.~Tang, H.~He, M.~Cheng, X.~Jin, M.~Zou, S.~Tang, X.~Ma, T.Y. Chen, J.W. Pan, Optica \textbf{11}(6), 883 (2024)

\bibitem{ge2025RN1181}
C.~Ge, L.~Zhou, J.~Lin, H.L. Yin, Q.~Zeng, Z.~Yuan, Quantum Science and Technology \textbf{10}(1), 015046 (2025)

\bibitem{zhou2025RN1272}
X.Y. Zhou, J.R. Hu, C.H. Zhang, Q.~Wang, Optics Letters \textbf{50}(2), 249 (2025)

\bibitem{xie2023RN1097}
Y.M. Xie, J.L. Bai, Y.S. Lu, C.X. Weng, H.L. Yin, Z.B. Chen, Physical Review Applied \textbf{19}(5), 054070 (2023)

\bibitem{zhou2025RN1275}
X.Y. Zhou, J.R. Hu, C.H. Zhang, Q.~Wang, Quantum Information Processing \textbf{24}(3), 98 (2025)

\bibitem{wang2023RN1051}
Z.H. Wang, R.~Wang, Z.Q. Yin, S.~Wang, F.Y. Lu, W.~Chen, D.Y. He, G.C. Guo, Z.F. Han, Communications Physics \textbf{6}(1), 265 (2023)

\bibitem{lu2024RN1183}
Z.~Lu, G.~Wang, C.~Li, Z.~Cao, Physical Review A \textbf{109}(1), 012401 (2024)

\bibitem{li2024RN1225}
Z.~Li, T.~Dou, Y.~Xie, W.~Kong, Y.~Liu, H.~Ma, J.~Tang, arXiv:2412.12593v2  (2024)

\bibitem{li2024RN1271}
Z.~Li, T.~Dou, M.~Cheng, Y.~Liu, J.~Tang, Optics Letters \textbf{49}(23), 6609 (2024)

\bibitem{liu2023RN1182}
X.~Liu, D.~Luo, Z.~Zhang, K.~Wei, Physical Review A \textbf{107}(6), 062613 (2023)

\bibitem{zhou2024RN1084}
X.Y. Zhou, J.R. Hu, J.J. Wang, Y.~Cao, C.H. Zhang, Q.~Wang, Optics Express \textbf{32}(10), 18366 (2024)

\bibitem{cao2015RN105}
Z.~Cao, Z.~Zhang, H.K. Lo, X.~Ma, New Journal of Physics \textbf{17}(5), 053014 (2015)

\bibitem{zhang2024RN1294}
C.M. Zhang, Z.~Wang, Y.D. Wu, J.R. Zhu, R.~Wang, H.W. Li, Physical Review A \textbf{109}(5), 052432 (2024)

\bibitem{gottesman2004RN266}
D.~Gottesman, H.K. Lo, N.~Lütkenhaus, J.~Preskill, Quantum Information and Computation \textbf{4}(5), 325 (2004)

\bibitem{zhang2017RN87}
Z.~Zhang, Q.~Zhao, M.~Razavi, X.~Ma, Physical Review A \textbf{95}(1), 012333 (2017)

\bibitem{curty2014RN100}
M.~Curty, F.~Xu, W.~Cui, C.C. Lim, K.~Tamaki, H.K. Lo, Nature Communications \textbf{5}(1), 3732 (2014)

\bibitem{lim2014RN93}
C.C.W. Lim, M.~Curty, N.~Walenta, F.~Xu, H.~Zbinden, Physical Review A \textbf{89}(2), 022307 (2014)

\end{thebibliography}

\end{document}